\documentclass[aps,prl,reprint,superscriptaddress,longbibliography,nofootinbib,floatfix,raggedbottom]{revtex4-2}
\usepackage[T1]{fontenc}
\usepackage{amsmath,amssymb,bm,graphicx}
\usepackage{microtype}
\usepackage[colorlinks=true,allcolors=blue]{hyperref}
\hypersetup{pdftitle={Strong Collective Dynamics behind a Nearly Free Single-Particle Spectrum}}
\newcommand{\tr}{\operatorname{tr}}
\newcommand{\im}{\operatorname{Im}}

\newcommand{\spec}{\mathcal A}
\newcommand{\hc}{\mathrm{H.c.}}
\begin{document}
\title{Strong Collective Dynamics Hidden behind a Nearly Free Single-Particle Spectrum}
\author{Zhixuan Zhao}
\author{Sicheng He}
\author{Hanxiang Di}
\author{Nayun Chen}
\author{Jiabai Xiong}
\author{Guyixue Gao}
\author{Yanling Wu}
\author{Jun Li}
\email{ljcj007@ysu.edu.cn}
\affiliation{State Key Laboratory of Metastable Materials Science and Technology,
	Hebei Key Laboratory of Microstructural Material Physics,
	School of Science, Yanshan University, Qinhuangdao, 066004, China}
\date{\today}
\begin{abstract}
Single-particle spectroscopy has long served as a criterion for the strength of electronic correlations: strong collective motion is expected to shift and broaden spectral peaks and to renormalize the dispersion. Here we show that this correspondence can fail on the same microscopic orbital. The mechanism relies on local interference between two resonant side orbitals: one suppresses low-energy real transitions into the interacting environment, while the other suppresses spectral-weight backflow to the source orbital; virtual mixing nevertheless transfers the environmental spin polarization to the source. We prove a rigorous relative spectral bound that is uniform in environment size and interaction strength and holds for arbitrary stationary states; the protection becomes stronger as the relative frequency resolution improves. In a fully quantum two-cell device, a collective mode softens by 60\%, whereas the particle intensity at the free peak changes by only 3.6 parts per million. Quantum Monte Carlo shows that the transferred spin response grows with spatial correlations and is quantitatively predicted by the free-band polarization. These results recast the meaning of a simple particle spectrum and provide a design principle for quantum probes that combine predictable particle spectra with strong collective sensitivity.
\end{abstract}
\maketitle

Single-particle spectroscopy is a central tool for identifying strong electronic correlations. The conventional view is that strong collective motion should leave a strong imprint on the single-particle spectra of the electrons involved. This expectation has a concrete physical basis: coupling to spin fluctuations and lattice vibrations changes the electron self-energy, shifting and broadening spectral peaks and renormalizing the dispersion~\cite{eschrig2003,lanzara2001}. Photoemission and tunneling spectroscopy exploit these changes to identify collective modes and their coupling to electrons~\cite{damascelli2003,sobota2021,lee2006}. Within this picture, a nearly free low-energy spectrum suggests weak collective effects on the measured electrons.

We demonstrate that this conventional connection can be broken on the same microscopic orbital. Its collective motion can be strongly reshaped by interactions while its complete low-energy particle spectrum approaches a fixed free reference. We prove a relative spectral bound, uniform in environment size and interaction strength and valid in every stationary state of the specified environment. The contrast is striking in a fully quantum two-cell device: a collective mode softens by $60\%$, yet the particle intensity at the free peak changes by only $3.6$ parts per million at the same frequency resolution. In two dimensions, the protected orbital also carries a spin response that grows with spatial correlations.

The result changes what a simple low-energy spectrum tells us. The measured orbital can itself carry strongly renormalized collective motion even when its particle spectrum looks nearly free. This separation gives quantum sensing a new design principle: a probe can retain a predictable particle spectrum while detecting the many-body dynamics around it.

We realize this separation through local interference. Interference can cancel electronic propagation~\cite{fano1961,miroshnichenko2010}, alter impurity screening~\cite{diasdasilva2006,diasdasilva2008prb,fritz2004}, and suppress scattering from critical fluctuations~\cite{wang2025,paul2017}; spectral pinning also preserves particular features under strong interactions~\cite{langreth1966,peters2013,eickhoff2024}. Here two resonant side orbitals protect the complete low-energy particle spectrum. They restrict real transitions into the interacting environment and suppress their effect on the source. At the same time, virtual mixing with higher bands redistributes the spin of occupied states and transfers collective polarization to the protected orbital. The local geometry gives us control over both responses.

\begin{figure}[tbp]
\includegraphics[width=\columnwidth]{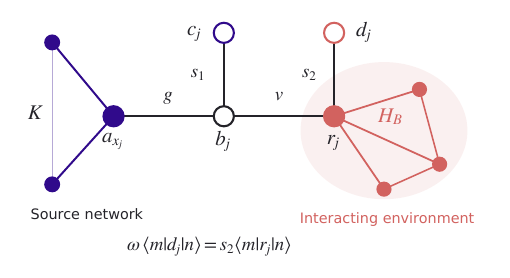}
\caption{Local geometry separates spectral protection from collective participation. Source orbitals $a_x$ connect through $b$ to the interacting target $r$. The resonant side orbitals $c,d$ are noninteracting. Orbital $d$ restricts low-energy transitions on $r$; $c$ suppresses their effect on the source. Virtual mixing with higher bands still transfers spin polarization to $a$. Repeating the contact produces a metal.}
\label{fig:geometry}
\end{figure}

\textit{Protecting the complete particle spectrum.---}We begin with the particle spectrum and ask how an orbital can remain spectrally predictable while connected to an interacting environment. Figure~\ref{fig:geometry} places two side orbitals along the path to that environment. We call the measured orbitals $a_x$ the source. The Hamiltonian, with real contact hoppings and $\hbar=1$, is
\begin{align}
H={}&\sum_{xy}K_{xy}a_x^\dagger a_y+H_B(r,\mathcal B)\nonumber\\
 &+\sum_j\bigl(g a_{x_j}^\dagger b_j+s_1c_j^\dagger b_j\nonumber\\
 &\hspace{17mm}+v b_j^\dagger r_j+s_2d_j^\dagger r_j+\hc\bigr).
\label{eq:model}
\end{align}
The orbital symbols are fermion annihilation operators, with spin labels suppressed; H.c. denotes the Hermitian conjugate. The matrix $K$ contains source hopping and onsite energies, and $x_j$ is the source site at contact $j$. All interactions reside in $H_B$, acting on the targets $r$ and other environment degrees of freedom $\mathcal B$. They may be arbitrarily strong and must conserve fermion parity. The source, connectors, and side orbitals remain quadratic, with \mbox{$s_1s_2\ne0$}. The common side-orbital energy defines frequency zero and, in the metal, coincides with the chemical potential. Unequal contacts and matrix-valued couplings are treated in the Supplemental Material~\cite{supplement}.

We compare the interacting device with its fixed free reference: the same connected network with fermion interactions removed and all one-body parameters retained. A particle spectrum combines the transitions produced by adding and removing an electron. Their sum is independent of occupation in the free reference. For a stationary state $\rho$, $[\rho,H]=0$, it is given by
\begin{align}
 G_{xy}^\rho(z)&=-i\int_0^\infty dt\,e^{izt}\tr\rho\{a_x(t),a_y^\dagger\},\nonumber\\
 \spec^\rho(z)&=-\im G^\rho(z)/\pi,\qquad z=\omega+i\eta.
\label{eq:spectrum}
\end{align}
Here $a_x(t)=e^{iHt}a_xe^{-iHt}$, $\im G=(G-G^\dagger)/(2i)$, and $\eta>0$ is the Lorentzian half-width. The diagonal entry $\spec^\rho_{xx}$ is the local intensity; $u^\dagger\spec^\rho u$ is the intensity of a normalized spatial probe $u$. These spectral weights are positive and integrate to the identity~\cite{lehmann1954,zubarev1960}. We denote the free spectrum by $\spec_0$.

The target-side orbital $d$ supplies the essential restriction. Its exact equation of motion, $i\partial_t d_j=s_2r_j$, relates every transition of the full interacting system:
\begin{equation}
 \langle m|r_j|n\rangle=\frac{E_n-E_m}{s_2}\langle m|d_j|n\rangle,
 \label{eq:physical}
\end{equation}
where $|n\rangle$ has energy $E_n$. The target amplitude of every low-energy transition is therefore small, regardless of its many-body origin. Since $d$ has unit total spectral weight, the target weight below energy $E$ is bounded by $E^2/s_2^2$. The second side orbital, $c$, suppresses the propagation of this weight back to the source. Together, the two orbitals give the bound
\begin{align}
 \delta(z)&=\frac{|z|^4}{\varepsilon_g^2\eta^2},\qquad
 \varepsilon_g=|s_1s_2/v|,\nonumber\\
 (1-\delta)\spec_0(z)&\preceq\spec^\rho(z)\preceq\frac{\spec_0(z)}{1-\delta},
 \qquad\delta<1.
 \label{eq:main}
\end{align}
The ordering means that the inequality holds for every probe $u$. Every intensity therefore differs from its reference by at most the fraction $\mathcal R=\delta/(1-\delta)$. This guarantee is set entirely by the contact: it is unchanged when the environment grows or its interactions become stronger. Every stationary state is covered, including infinite temperature. The comparison includes the complete addition--removal spectrum and all resolution tails.

The protection becomes stronger as the measurement resolves the excitation more finely. For an excitation at $|\omega|=E\ll\varepsilon_g$, choose
\begin{equation}
 E^2/\varepsilon_g\ll\eta\ll E.
 \label{eq:window}
\end{equation}
For example, $\eta=\varepsilon_g(E/\varepsilon_g)^{3/2}$ makes both the relative error $\mathcal R$ and the relative linewidth $\eta/E$ vanish. The source becomes spectrally indistinguishable from its fixed free reference at an increasingly fine relative resolution. The lower bound on $\eta$ resolves the physical role of the contact: thermal resonances can split a source peak on the scale $E^2/\varepsilon_g$, and the window averages over this fine splitting while resolving the excitation itself~\cite{supplement}.

The proof uses the side-orbital equations at both ends of the Green function and normalizes their effect by the free source intensity. It therefore controls interaction-generated transitions as well as the original free modes. The second side orbital is decisive: with only $d$, an allowed narrow resonance can retain a finite fractional distortion at low energy. The Supplemental Material derives the bound and its extension to other interference geometries~\cite{supplement}.

\begin{figure*}[t]
\includegraphics[width=\textwidth]{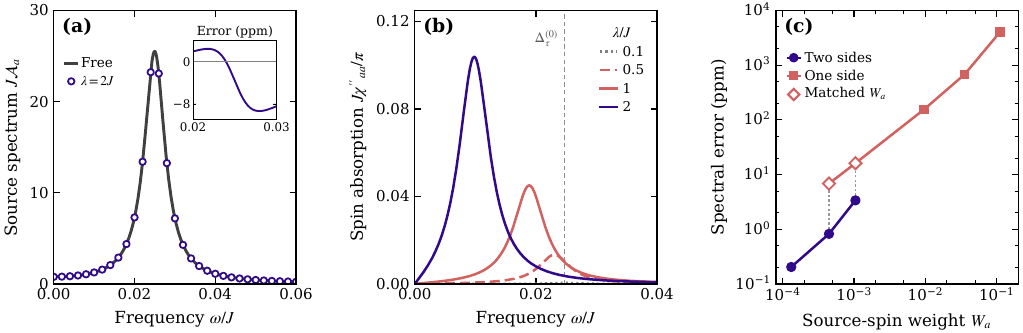}
\caption{A strongly softened collective mode with a nearly unchanged particle spectrum. The quantum two-cell device has ten spinful orbitals, two Ising spins, contact hoppings $J$, source hopping $0.05J$, and $J_\tau=h_\tau=0.02J$. (a) Full local source particle spectrum at $\lambda=2J$, with $\eta=0.003125J$; inset: $10^6(\mathcal A/\mathcal A_0-1)$ near the free peak. (b) Full source-spin absorption for $M_a=(M_{a_1}+M_{a_2})/\sqrt2$, at the same half-width. The dashed line marks the isolated Ising-pair gap. Both signs of frequency and all higher-state tails are included. (c) Particle-spectrum error at each geometry's free peak, with $\eta=0.005J$, versus collective weight $W_a=|\langle0|M_a|p\rangle|^2$. Here $|0\rangle$ is the sector ground state and $|p\rangle$ the collective excited state. Filled markers use $\lambda/J=0.5,1,2$; open diamonds match the one-side weight to the two-side values at $\lambda/J=1,2$.}
\label{fig:collective}
\end{figure*}

\textit{A protected orbital that carries collective motion.---}The particle bound describes adding or removing an electron. We now turn to the electrons already present and ask how their spins move together. Their wavefunctions can redistribute spin among orbitals, giving the source a share of the environment's collective dynamics. To calculate this response, we couple quantum Ising spins to the target spin density:
\begin{align}
 H_B={}&-J_\tau\sum_{\langle ij\rangle}\tau_i^z\tau_j^z
 -h_\tau\sum_i\tau_i^x\nonumber\\
 &+\lambda\sum_i\tau_i^z M_{r_i}.
 \label{eq:spin}
\end{align}
Here $\tau^{x,z}$ are Pauli operators and $M_\alpha=n_{\alpha\uparrow}-n_{\alpha\downarrow}$. The exchange $J_\tau$ aligns neighboring spins; the transverse field $h_\tau$ drives quantum fluctuations. Each Ising spin supplies opposite target potentials to the two fermion species. Reversing the Ising spins and exchanging the species is an exact symmetry. Setting $\lambda=0$ gives the free particle reference.

In a periodic array, take all contact hoppings equal to $J$ and the source dispersion to be $J[0.05-0.10(\cos k_x+\cos k_y)]$, with lattice spacing one. One band crosses the Fermi level, while the other four remain more than $0.5J$ away. The low-band target amplitude vanishes as $u_r(\bm k)=O(e_{\bm k}^2)$, where $e_{\bm k}$ is the band energy relative to the Fermi level. Real low-energy scattering is consequently weak. A slowly varying target field also mixes occupied states with the other bands, changing their orbital composition. This virtual mixing transfers spin without exciting a particle across the band gap.

A free-band response tells us how much spin is transferred to each orbital. Apply a small uniform potential $m$ to the target orbitals of one species and define $\kappa_\alpha=\partial_m\langle n_\alpha\rangle|_{m=0}$ at zero temperature and fixed chemical potential. A slowly varying Ising polarization $\phi$ then induces the orbital spin $b_\alpha\phi$, with $b_\alpha=2\lambda\kappa_\alpha+O(\lambda^3)$. The factor two comes from the opposite potentials of the two species. The interband gap makes this transfer local on the long scales of the collective motion.

Correlations between Ising spins are carried to the induced orbital spins across the array. Near an Ising transition, their leading singular response takes the form
\begin{equation}
 \chi_{\alpha\beta}^{\mathrm{sing}}(q,\omega)
   =b_\alpha b_\beta\chi_{\phi\phi}(q,\omega),
 \label{eq:collective}
\end{equation}
where $\chi_{\alpha\beta}$ is the spin response on orbital $\alpha$ to a field on orbital $\beta$, at wave number $q$ and frequency $\omega$. The source coefficient is nonzero: $\kappa_a=-0.00953/J$. The collective mode therefore appears in the response of the protected orbital itself, with a calculable spectral weight. The induced spins compensate within a cell, $\sum_\alpha b_\alpha=0$, while each orbital follows the collective fluctuation across cells~\cite{supplement}. Collective participation remains visible locally even when the net induced spin in a cell cancels.

\textit{A large collective change with a nearly free particle spectrum.---}The clearest test is to measure both responses on the same orbital in the same quantum state. We use a two-cell device, retaining all fermion bands and the quantum Ising dynamics. Its lowest state in the half-filled sector gives access to the full particle and spin spectra at strong coupling. At $\lambda=2J$, the collective gap falls from the isolated pair's $0.0247J$ to $0.0098J$, a $60\%$ decrease. The source spin absorbs at this shifted frequency [Fig.~\ref{fig:collective}(b)]. At the same resolution, $\eta=0.003125J$, its particle intensity at the free peak changes by only $3.6$ parts per million [Fig.~\ref{fig:collective}(a)]. The two signals come from the same orbitals in the same state. The collective excitation shifts dramatically, while the particle intensity remains nearly fixed. The full spectra, including higher-energy tails, show this separation directly~\cite{supplement}.

We next ask how spectral protection affects collective sensitivity. We compare with a one-side contact $a-r-d$, matching the free source weight and low-band velocity. We then tune its coupling to match the integrated source-spin peak weight $W_a$. At equal $W_a$, its particle-spectrum error is $4.8$--$8.3$ times larger [Fig.~\ref{fig:collective}(c)]. The second side orbital therefore gives a cleaner particle reference at the same collective signal. This comparison fixes the collective peak weight while allowing its frequency to differ.

\begin{figure*}[t]
\includegraphics[width=\textwidth]{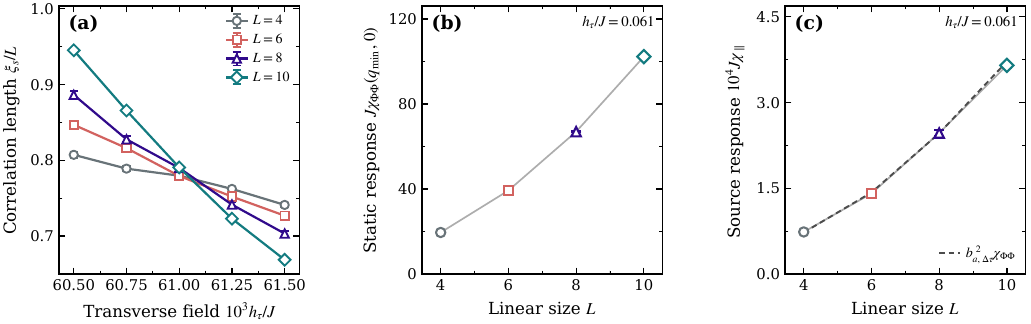}
\caption{Collective correlations grow on the spectrally protected orbital. Quantum Monte Carlo uses $L=4,6,8,10$, $\lambda=0.1J$, $J_\tau=0.02J$, $\beta J=50L$, and $\Delta\tau J=0.0625$. (a) Correlation length divided by size across the ordering region. (b) Static Ising response at $q_{\min}=2\pi/L$ and $h_\tau=0.061J$. (c) Source component $\chi_{\parallel}=|\chi_{a\Phi}|^2/\chi_{\Phi\Phi}$ at the same momentum and field. The dashed prediction $b_{a,\Delta\tau}^2\chi_{\Phi\Phi}$ uses an independent free-band coefficient at the same size, temperature, momentum, and time step, together with the paired Ising response. Error bars are one joint block-bootstrap standard error; solid lines join measured points. Estimators and sampling parameters are given in the Supplemental Material.}
\label{fig:scaling}
\end{figure*}

\textit{Collective correlations across a metal.---}In a metal, spin correlations can extend across many cells and connect the protected orbitals throughout the array. Their evolution depends on feedback from itinerant electrons. An ordinary metal damps spin fluctuations by creating low-energy particle--hole pairs, giving absorption proportional to $\omega/q$~\cite{hertz1976,millis1993}. Here the particle and hole each carry target probability of order $E^4$. Their product supplies eight extra energy powers: absorption becomes $\omega^9/q$ for $\omega\ll v_Fq$, with Fermi velocity $v_F$. 

To determine the fate of the transition, we also account for generated couplings, including those to the Ising energy density~\cite{nandkishore2012}. The target-amplitude factors survive band elimination and keep the induced spin interactions short ranged in a convergent weak-coupling expansion~\cite{supplement}. Under smooth flow within the stable basin of the short-range Ising fixed point, these local corrections shift the ordering field by $O(\lambda^2)$. Retuning $h_\tau$ then gives a $2+1$-dimensional Ising transition and the singular orbital response in Eq.~\eqref{eq:collective}~\cite{wilson1971,wegner1972,fisher1998,kos2016}. Throughout this collective evolution, Eq.~\eqref{eq:main} continues to protect the particle spectrum.

Determinant quantum Monte Carlo measures this collective participation in the full five-orbital model~\cite{wolff1989,schattner2016}. Periodic $L\times L$ lattices have $\lambda=0.1J$, $J_\tau=0.02J$, inverse temperature $\beta J=50L$, and time step $\Delta\tau J=0.0625$. Crossings of the correlation-length ratio $\xi_s/L$ locate the ordering region [Fig.~\ref{fig:scaling}(a)]. We follow the transferred response at fixed field $h_\tau=0.061J$.

A field on the Ising mode $\Phi$ produces a response $\chi_{\Phi\Phi}$ in the Ising spins and a response $\chi_{a\Phi}$ at the source. Their ratio measures how much source polarization follows each unit of Ising polarization. Static thermal response defines a positive inner product. We use it to project the source spin onto $\Phi$ and obtain the exact decomposition
\begin{align}
 \chi_{aa}&=\chi_{\parallel}+\chi_{\perp},\qquad \chi_{\perp}\ge0,\nonumber\\
 \chi_{\parallel}&=\frac{|\chi_{a\Phi}|^2}{\chi_{\Phi\Phi}}.
 \label{eq:qmc-participation}
\end{align}
The component $\chi_{\parallel}$ is the part of the source's own response that follows the Ising spins. Here $\Phi$ is their Fourier mode normalized by $1/L$, and all responses use the same momentum and static limit. At $q_{\min}=2\pi/L$, $J\chi_{\Phi\Phi}$ rises from $19.53(7)$ to $102.26(23)$ between $L=4$ and $10$. The source component grows nearly fivefold, from $J\chi_{\parallel}=7.35(19)\times10^{-5}$ to $3.65(5)\times10^{-4}$. Its magnitude follows the independently calculated free-band polarization [Fig.~\ref{fig:scaling}(b,c)]. At $\lambda=0$ the mixed response vanishes. Coupling to the environment therefore makes the spectrally protected source share its growing spatial correlations.

\textit{Measuring the separation.---}Two measurements on the same orbital make the separation visible. Particle addition and removal reveal its spectrum, while a weak spin-dependent field probes its collective response. A uniform field probes collective spin absorption below the interband gap, where the conserved free low-band occupations cannot absorb. At small nonzero momentum, probing above the free particle--hole continuum similarly exposes the transferred collective signal, with weight proportional to $b_a^2$~\cite{supplement}. A patterned field resolves spatial correlations; a single-site field sums over momenta.

Particle spectroscopy combines addition and removal signals or uses a number-conserving Ramsey sequence~\cite{bishop2025}, with resolution in Eq.~\eqref{eq:window}. Controlled tunneling, onsite interactions, and resolved detection in fermionic tweezer arrays provide ingredients for a finite-device realization~\cite{spar2022,yan2022,jain2026}. The Supplemental Material derives both readouts and their common window set by temperature, size, contact errors, and weak residual source interactions. Detailed preparation and acquisition protocols accompany the data.

A single orbital can carry strong collective dynamics behind a nearly free low-energy particle spectrum. Local geometry protects that spectrum through a rigorous relative bound and transmits a calculable collective spin response. In the quantum device and extended array, the protected orbital reveals both mode softening and growing spatial correlations. These results give a new meaning to a simple particle spectrum and a design principle for quantum probes of collective matter.

\emph{Acknowledgments}---
This work is funded by Natural Science Foundation of Hebei Province (Grant No. A2026203033), Beijing National Laboratory for Condensed Matter Physics (Grant No. 2024BNLCMPKF020).  
\bibliography{references}
\end{document}


\maketitle
\vspace{-2em}

The central question is how much collective physics a simple particle spectrum can hide. The Letter answers it with two results on the same microscopic orbital: a relative bound on its complete particle spectrum, and a collective spin response whose dynamics can change strongly. This supplement develops the argument in that order and supplies the calculations behind the figures.

Sections~\ref{sec:model} and \ref{sec:proof} define the measured spectrum and prove its protection. Section~\ref{sec:collective} calculates the spin transferred by virtual polarization. Sections~\ref{sec:quantum-device} and \ref{sec:qmc-scaling} then follow the particle and spin responses in the quantum device and the extended lattice. The effect of electronic feedback on a critical transition is developed in Sec.~\ref{sec:collective-feedback}.

Sections~\ref{sec:collective-measurement} and \ref{sec:implementation} connect these results to measurements: which probes access the two responses, and which energy scales leave a common working window. Each calculation below supports one of these steps.

A separate companion, \emph{Extended Calculations and Measurement Protocols}, collects the broader contact-design framework, the disordered-chain application, high-frequency spectral moments, finite-bath examples, and detailed acquisition protocols. It also contains the full resolution scans and interval-arithmetic implementation behind the bounds quoted here. These calculations and their numerical inputs accompany the source package.

We use $\hbar=1$ except when converting to laboratory times. A spectral function includes both particle addition and removal unless a channel is specified. The free reference retains every hopping amplitude and onsite energy when the interaction is removed.

\section{The full spectral measure and the model}
\label{sec:model}

Adding an electron and rotating its spin ask different questions of a many-body state. Here we first define the particle measurement. Its spectrum must include every allowed addition and removal transition, because an interaction can create new lines as well as move existing ones. The positive spectral measure below keeps all of this weight and gives the sum rule on which the proof rests.

Let $f_i$ be canonical fermion annihilation operators and $\rho$ a stationary density matrix, $[\rho,H]=0$. We evolve the operators with the full Hamiltonian, $f_i(t)=e^{\ii Ht}f_ie^{-\ii Ht}$. The retarded normal Green matrix and its Lorentzian-broadened spectrum are
\begin{align}
 G_{ij}^{\rho}(z)&=-\ii\int_0^\infty\dd t\,e^{\ii zt}\Tr\rho\{f_i(t),f_j^\dagger\},
 &z&=\omega+\ii\eta,\quad\eta>0,\label{eq:Gdef}\\
 \A^\rho(z)&=-\frac{G^\rho(z)-G^\rho(z)^\dagger}{2\pi\ii}.
 \label{eq:Adef}
\end{align}
We use the matrix imaginary part $\ImH G=(G-G^\dagger)/(2\ii)$ throughout, including spatial off-diagonal responses. Choose common eigenstates $H|n\rangle=E_n|n\rangle$ and $\rho|n\rangle=p_n|n\rangle$. With transition amplitudes $x_{mn,i}=\langle m|f_i|n\rangle$, the Lehmann representation is~\cite{lehmann1954,zubarev1960}
\begin{equation}
 G^\rho(z)=\int_{\mathbb R}\frac{\dd M^\rho(\epsilon)}{z-\epsilon},\qquad
 \dd M^\rho(\epsilon)=\sum_{m,n}(p_m+p_n)x_{mn}x_{mn}^\dagger
 \delta_{E_n-E_m}(\dd\epsilon).
 \label{eq:measure}
\end{equation}
Here $\delta_{E_n-E_m}$ is a unit point mass at the transition energy $E_n-E_m$. We write $\dd\mu_x^\rho=(\dd M^\rho)_{xx}$ for the local spectral measure and $A_x^\rho(\omega;\eta)=(\A^\rho(\omega+\ii\eta))_{xx}$ for its broadened intensity. In the single-source resonance example below, $A_U$ and $A_0$ denote this quantity with and without the interaction.

Thus $\dd M^\rho\succeq0$ and $M^\rho(\mathbb R)=I$ by the canonical anticommutation relations. A normalized spatial probe $f(u)=\sum_i\overline u_i f_i$ has a positive measure $u^\dagger\dd M^\rho u$ of total mass one. This representation holds for every stationary state, including thermal mixtures and individual energy eigenstates.

We next specify which parts of the device can interact. The source vector $a$ has $N$ components. There are $M$ first connectors $b$, first side orbitals $c$, interacting targets $r$, and second side orbitals $d$. The numbers of branches and source sites are independent. In block notation the Hamiltonian is
\begin{align}
 H_\alpha&=C^\dagger h_C C+b^\dagger V r+r^\dagger V^\dagger b
       +r^\dagger S_2d+d^\dagger S_2^\dagger r+H_{B,\alpha}(r,\mathcal B),
       \qquad C=(a,b,c)^T,\label{eq:H}\\
 h_C&=\begin{pmatrix}K&T&0\\T^\dagger&D_b&S_1\\0&S_1^\dagger&0\end{pmatrix}.
 \label{eq:hC}
\end{align}
Here $K=K^\dagger$ is any source hopping matrix, $D_b=D_b^\dagger$, $T$ is an $N\times M$ source-to-connector matrix, and $S_1,S_2$ are invertible $M\times M$ matrices. The zero blocks specify the common side-orbital resonance energy. The self-adjoint, parity-even $H_{B,\alpha}$ is supported exclusively on $r$ and additional bath orbitals $\mathcal B$. It can contain target hopping, interactions of arbitrary strength, and pairing confined to the bath. The external couplings displayed in Eq.~\eqref{eq:H} conserve particle number. The finite-dimensional spectral argument applies to arbitrary matrices; finite-range matrices give a local network.

The index $\alpha$ labels two allowed environments with the same external geometry. Each is evaluated in its own stationary state. For the comparison in the Letter, $\alpha=0$ is the predetermined quadratic reference and $\alpha=1$ includes the interaction. A quadratic anticommutator spectrum is independent of the reference state. The scalar notation of the Letter is recovered for $N=M$, $T=gI$, $D_b=\epsilon_bI$, $S_1=s_1I$, $S_2=s_2I$, and $V=vI$, with real couplings.

\subsection{Addition, removal, and thermal occupations}
\label{sec:channels}

A removal measurement requires an occupied initial state, whereas an addition measurement requires an available final state. For example, photoemission measures the occupied part of the spectral function, with occupation factors and instrumental broadening entering separately~\cite{damascelli2003}. The two terms in Eq.~\eqref{eq:measure} make this distinction explicit: $p_n$ weights removal and $p_m$ weights addition. Their sum has unit total weight for a normalized probe. Its frequency distribution can still change with the interacting state.

For a number-conserving Hamiltonian, grand-canonical equilibrium has $\rho\propto e^{-\beta(H-\mu N)}$, with inverse temperature $\beta$, chemical potential $\mu$, and total particle number $N$. The unbroadened measures obey
\begin{equation}
 \dd M_-(\epsilon)=f_\beta(\epsilon-\mu)\dd M(\epsilon),\qquad
 \dd M_+(\epsilon)=[1-f_\beta(\epsilon-\mu)]\dd M(\epsilon),\quad
 f_\beta(x)=\frac{1}{e^{\beta x}+1}.
 \label{eq:thermalchannels}
\end{equation}
Here $-$ denotes removal, $+$ addition, and frequencies are generated by $H$. With Lorentzian resolution, the removal signal is instead
\begin{equation}
 \A_-(\omega;\eta)=\int_{\mathbb R}
 \frac{\eta\,f_\beta(\epsilon-\mu)}{\pi[(\omega-\epsilon)^2+\eta^2]}
 \dd M(\epsilon).
 \label{eq:broadenedchannel}
\end{equation}
The Fermi factor remains inside the integral. Thus low temperature and a probe frequency above $\mu$ suppress removal near that frequency, but occupied lines elsewhere can still contribute through the resolution tails.

A single channel inherits a quantitative comparison when its complement is small in both spectra. For a fixed source probe satisfying the total-spectrum bound, let $A$ and $A_0$ be the total broadened intensities. Choose the same addition or removal channel in both spectra. If the other channel carries at most a fraction $\epsilon$ of each and $\R<1$, the dominant channel obeys
\begin{equation}
 (1-\R)(1-\epsilon)\le
 \frac{A_{\mathrm{dom}}}{A_{0,\mathrm{dom}}}
 \le\frac{1+\R}{1-\epsilon},\qquad 0\le\epsilon<1.
 \label{eq:dominantchannel}
\end{equation}
This follows by writing $A_{\mathrm{dom}}=A-A_{\mathrm{comp}}$ in each spectrum. The required dominance concerns the measured, broadened intensities, so it includes both thermal occupations and resolution tails. The sum measured by the empty/filled-probe protocol obeys the main bound directly in every stationary state. The fixed-particle-number mixtures used in the numerical examples require no grand-canonical assumption.

\section{Relative protection of the complete source spectrum}
\label{sec:proof}

The spectral guarantee follows from two exact properties: a canonical orbital has unit total spectral weight, and a noninteracting side orbital obeys a linear equation of motion. These properties let us control the source spectrum for the interacting environment as a whole. We first use the sum rule to bound any environment at a fixed resolution. The equation for $d$ then restricts transitions on the target, and the equation for $c$ restricts their propagation to the source. Dividing by the source's own free intensity turns these restrictions into the relative bound used in the Letter.

\subsection{A bound shared by all canonical Green matrices}

A narrow resonance can make a Green function large, but the unit spectral weight limits how large it can become at fixed resolution $\eta$. To obtain a bound shared by all environments, write any matrix $Q(z)$ with the positive, normalized representation \eqref{eq:measure} as
\begin{equation}
 Q(z)=J^\dagger(z-L)^{-1}J,\qquad J^\dagger J=I,
 \label{eq:dilation}
\end{equation}
where $L$ is self-adjoint and $J$ embeds the probe space without changing vector lengths. This expresses the Green matrix as a projection of a resolvent. One construction takes $L$ to multiply by the spectral variable in the $L^2$ space of the measure, and $J$ maps vectors to constant functions. Consequently
\begin{equation}
 I-2\ii\eta Q(z)=J^\dagger(\overline z-L)(z-L)^{-1}J,
 \qquad \|I-2\ii\eta Q(z)\|\le1.
 \label{eq:ball}
\end{equation}
The operator between $J^\dagger$ and $J$ is unitary. All such Green matrices therefore lie in the same norm ball,
\begin{equation}
 \left\|Q(z)+\frac{\ii I}{2\eta}\right\|\le\frac1{2\eta}.
 \label{eq:ball2}
\end{equation}
For any two canonical Green matrices, even from different Hamiltonians and states, this gives
\begin{equation}
 \|Q_1-Q_0\|\le\eta^{-1},\qquad
 \|z^2Q-zI\|\le |z|^2/\eta.
 \label{eq:canonical}
\end{equation}
For the second inequality, multiply the ball in Eq.~\eqref{eq:ball2} by $z^2$ and subtract $zI$. The new radius is $|z|^2/(2\eta)$ and its center is $-\ii|z|^2I/(2\eta)$. The distance from zero is therefore at most twice that radius. These bounds require only spectral positivity and unit total weight; no information about the interacting excitation energies enters them.

\subsection{The second side orbital constrains every transition}

The exact equation $\ii\partial_t d=S_2^\dagger r$ implies, between exact eigenstates at frequency $\epsilon=E_n-E_m$,
\begin{equation}
 S_2^\dagger r_{mn}=\epsilon d_{mn},\qquad
 S_2^\dagger\dd M_{rr}^\rho(\epsilon)S_2
       =\epsilon^2\dd M_{dd}^\rho(\epsilon).
 \label{eq:actualtransition}
\end{equation}
In the scalar case,
\begin{equation}
 M_{rr}^\rho([-E,E])\preceq \frac{E^2}{s_2^2}I.
 \label{eq:targetweight}
\end{equation}
The frequency is a difference of energies of the \emph{interacting} Hamiltonian. Equation~\eqref{eq:targetweight} constrains both the original low-energy excitations and other transitions that interactions bring into the same frequency interval.

We now turn this transition-by-transition restriction into a bound on the Green function at finite resolution. Let $F_\alpha=G_{rr}^{(\alpha)}$ and $Q_\alpha=G_{dd}^{(\alpha)}$, both in the full coupled system. Insert Eq.~\eqref{eq:actualtransition} into the spectral integral and use
$\epsilon^2/(z-\epsilon)=z^2/(z-\epsilon)-z-\epsilon$.
The $d$ measure has zeroth moment $I$ and first moment zero: its free equation of motion contains $r$, whose anticommutator with $d^\dagger$ vanishes. Hence
\begin{equation}
 S_2^\dagger F_\alpha S_2=z^2Q_\alpha-zI,
 \qquad
 S_2^\dagger(F_1-F_0)S_2=z^2(Q_1-Q_0).
 \label{eq:leafGF}
\end{equation}
The common contact term $-zI$ cancels in the comparison.

\subsection{Propagation to the source and normalization by its signal}

Put $g_C(z)=(z-h_C)^{-1}$. This describes the free block $C=(a,b,c)$ with the bridge to $r$ removed. The reference $G^{(0)}$ includes that bridge and the full quadratic environment. Interactions act only outside $C$, so its equations of motion remain linear. Applying them at both ends of the Green function gives the exact Fano identity~\cite{diasdasilva2008}
\begin{equation}
 G_{aa}^{(\alpha)}-g_{C,aa}
   =g_{C,ab}VF_\alpha V^\dagger g_{C,ba}.
 \label{eq:fano}
\end{equation}
The bath Green matrix in this expression is that of the \emph{full} Hamiltonian, so Eq.~\eqref{eq:fano} already includes all repeated scattering and feedback.

Let $P_a,P_b,P_c$ select the source, connector, and side-orbital coordinates in $C$. For a source vector $u$, $\widetilde u=P_a^\dagger u$ embeds it in this larger space. The $c$ row of $(z-h_C)g_C\widetilde u=\widetilde u$ gives
\begin{equation}
 P_bg_C(z)\widetilde u=z(S_1^\dagger)^{-1}P_cg_C(z)\widetilde u.
 \label{eq:firstleaf}
\end{equation}
Apply the same relation at $\overline z$ to the other external leg. Define
\begin{equation}
 B=S_2^{-1}V^\dagger(S_1^\dagger)^{-1},\quad
 \kappa=\|B\|^2,\quad
 \delta(z)=\kappa\frac{|z|^4}{\eta^2}.
 \label{eq:kappa}
\end{equation}
The target comparison contributes two powers of $z$ through Eq.~\eqref{eq:leafGF}. Each propagation leg contributes one more through Eq.~\eqref{eq:firstleaf}. Bounding the remaining canonical Green-matrix difference by $1/\eta$ therefore gives
\begin{equation}
 |u^\dagger(G_{aa}^{(1)}-G_{aa}^{(0)})u|
 \le\frac{\kappa|z|^4}{\eta}
       \|g_C(\overline z)\widetilde u\|\,\|g_C(z)\widetilde u\|.
 \label{eq:twosides}
\end{equation}
The remaining propagator norms may be large near a source resonance. The free source intensity grows with exactly the same norms, so dividing by that intensity removes this apparent difficulty. Hermiticity of $h_C$ gives
\begin{equation}
 \|g_C(\overline z)\widetilde u\|=\|g_C(z)\widetilde u\|,\qquad
 -\ImH\bigl[u^\dagger g_{C,aa}(z)u\bigr]
       =\eta\|g_C(z)\widetilde u\|^2.
 \label{eq:ward}
\end{equation}
Thus
\begin{equation}
 |u^\dagger(G_{aa}^{(1)}-G_{aa}^{(0)})u|
 \le\delta(z)\bigl[-\ImH u^\dagger g_{C,aa}(z)u\bigr].
 \label{eq:relC}
\end{equation}
Using the second inequality of Eq.~\eqref{eq:canonical} instead gives the same estimate for $G_{aa}^{(\alpha)}-g_{C,aa}$. Hence, with $\A_C=-\ImH g_{C,aa}/\pi$,
\begin{equation}
 (1-\delta)\A_C\preceq\A_\alpha\preceq(1+\delta)\A_C.
 \label{eq:AC}
\end{equation}
The estimate is so far normalized to the disconnected block $C$. To express it relative to the complete free reference, use $\A_0\succeq(1-\delta)\A_C$ from Eq.~\eqref{eq:AC}. For $\delta<1$, replacing $\A_C$ by $\A_0/(1-\delta)$ in the spectral difference bound proves
\begin{equation}
 \boxed{(1-\R)\A_0\preceq\A_1\preceq(1+\R)\A_0,
 \qquad\R=\frac{\delta}{1-\delta}.}
 \label{eq:mainbound}
\end{equation}
The bound controls the positive spectral quadratic form of every source probe, from a single weakly responding site to a coherent spatial superposition. For uniform real couplings, $\kappa=v^2/(s_1^2s_2^2)$. The connector matrices alone determine this coefficient, giving the same guarantee for all source matrices, interaction strengths, stationary states, temperatures, and bath volumes.

The comparison is symmetric under interchange of the two environments at the same external geometry. Applying its upper bound in both directions gives
\begin{equation}
 (1-\delta)\A_0\preceq\A_1\preceq(1-\delta)^{-1}\A_0.
 \label{eq:positivebound}
\end{equation}
This multiplicative form retains a positive lower bound throughout $\delta<1$. The symmetric fractional allowance $\R$ is convenient when combining the spectral comparison with a measurement-error budget.

At $|\omega|=E$ and $\eta\ll E$, the error is asymptotically $\kappa E^4/\eta^2$. Thus
\begin{equation}
 \sqrt\kappa\,E^2\ll\eta\ll E
 \label{eq:window}
\end{equation}
makes both the fractional resolution $\eta/E$ and the relative error vanish. A dimensionally explicit choice is $\eta=E_*(E/E_*)^{1+\theta}$, where $0<\theta<1$ and $E_*$ is fixed. For the scalar geometry, $\kappa^{-1/2}=s_1s_2/|v|$ is the geometric energy in the Letter.

\subsection{What the relative bound adds to an exactly conserved mode}
\label{sec:spectral-consequences}

The proof controls every transition visible to the source. Two explicit examples explain why this is stronger than conserving one free mode, and why the resolving window has a lower scale proportional to the squared excitation energy.

\paragraph{The rest of the local spectral weight.}
\label{sec:dark}

Conserving one mode fixes its own spectral weight. A physical local probe also sees every other mode and transition with support on that orbital. A single-source contact makes this distinction explicit. Set the source and connector onsite energies to zero and take real couplings:
\begin{equation}
 H=g(a^\dagger b+\hc)+s_1(c^\dagger b+\hc)
   +v(b^\dagger r+\hc)+s_2(r^\dagger d+\hc)+H_B(r,\mathcal B).
 \label{eq:toyH}
\end{equation}
Write $\Omega^2=g^2+s_1^2$, $Z=s_1^2/\Omega^2$, and introduce the canonical combinations
\begin{equation}
 D=\frac{s_1a-gc}{\Omega},\qquad B=\frac{ga+s_1c}{\Omega},\qquad
 a=\sqrt Z D+\frac g\Omega B.
 \label{eq:DB}
\end{equation}
The fermion $D$ is absent from $H$, whereas $B$ couples to $b$. Hence $D(t)=D$ and $\{D,B^\dagger(t)\}=0$ as operator identities, valid in correlated states as well as product states. The local spectral measure is therefore
\begin{equation}
 \dd\mu_a=Z\delta_0+\frac{g^2}{\Omega^2}\dd\mu_B.
 \label{eq:DBmeasure}
\end{equation}
The exact bright equations $\ii\partial_tB=\Omega b$ and $\ii\partial_tb=\Omega B+vr$ imply
$(\epsilon^2-\Omega^2)B_{mn}=\Omega v r_{mn}$. On $|\epsilon|<\Omega$ this gives
\begin{equation}
 \dd\mu_a(\epsilon)=Z\delta_0(\dd\epsilon)
 +\frac{g^2v^2}{(\Omega^2-\epsilon^2)^2}\dd\mu_r(\epsilon).
 \label{eq:extraweight}
\end{equation}
The second term is positive. An environmental zero-frequency transition can therefore add weight to the same local spectral line while the original dark fermion is exactly unchanged.

When $s_2\ne0$, Eq.~\eqref{eq:actualtransition} makes the extra term
\begin{equation}
 \dd\mu_a(\epsilon)=Z\delta_0(\dd\epsilon)
 +\frac{g^2v^2\epsilon^2}{s_2^2(\Omega^2-\epsilon^2)^2}\dd\mu_d(\epsilon),
 \qquad |\epsilon|<\Omega.
 \label{eq:twoleafmeasure}
\end{equation}
The canonical sum rule now gives, for $0<E<\Omega$,
\begin{equation}
 0\le\mu_a([-E,E])-Z
 \le\frac{g^2v^2E^2}{s_2^2(\Omega^2-E^2)^2}.
 \label{eq:toyweightbound}
\end{equation}
In particular, $\mu_a(\{0\})=Z$. Equations~\eqref{eq:extraweight} and \eqref{eq:twoleafmeasure} apply within the open interval $|\epsilon|<\Omega$, with any spectral atoms at $\pm\Omega$ left unrestricted.

An exact thermal example illustrates the additional weight. Let $g=s_1=v=\Delta=1$ and
\begin{equation}
 H_B=\Delta n_r+U(n_r-\tfrac12)(n_f-\tfrac12),\qquad U=2,
 \label{eq:statictoy}
\end{equation}
where the occupation of the immobile fermion $f$ is conserved. In its two sectors the target onsite energies are $m=0,2$. With $s_2=0$, the source zero-line residues are respectively $2/3$ and $1/2$. At infinite temperature their equal mixture is $7/12$. With $s_2=1$, the zero-frequency $d$ equation forces $r=0$, and both sector residues are $1/2$. For each graph the free reference has residue $1/2$:
\begin{equation}
 \begin{array}{c|cc}
 &U=0&U=2\text{ at infinite temperature}\\\hline
 s_2=0&1/2&7/12\\
 s_2=1&1/2&1/2.
 \end{array}
 \label{eq:7over12}
\end{equation}
The reduction to quadratic sectors is exact because $f$ is conserved, as in Falicov--Kimball models~\cite{falicov1969,freericks2003}. Each graph is compared with its own complete free spectrum; their common free zero-line weight makes the additional interacting weight directly visible.

For a network, source hopping mixes the isolated-cell dark and bright combinations. The coupled transition equations extend the argument to this propagating setting. For uniform scalar couplings, the exact transition amplitudes satisfy
\begin{equation}
 \bigl[(\epsilon^2-\epsilon_b\epsilon-s_1^2)(\epsilon-K)-g^2\epsilon\bigr]a_{mn}
   =\frac{gv}{s_2}\epsilon^2d_{mn}.
 \label{eq:networktransition}
\end{equation}
Equation~\eqref{eq:mainbound} controls the resonant source propagation left implicit in this polynomial relation.

\subsection{A thermal resonance at the squared-energy scale}
\label{sec:resolution}

The squared-energy scale in Eq.~\eqref{eq:window} has a physical origin. An allowed interacting environment can split a source resonance on that scale. Resolving this splitting produces a finite change in the local spectrum. Consider six hopping-connected orbitals $(a,b,c,r,d,q)$ and one density-coupled conserved fermion $f$:
\begin{align}
 H_0(\epsilon_s)={}&\epsilon_s n_a+(\epsilon_s+\Delta)n_q
 +(a^\dagger b+c^\dagger b+r^\dagger d+r^\dagger q+\hc)
 +v(b^\dagger r+\hc),\label{eq:sharpH0}\\
 H_U(\epsilon_s)={}&H_0(\epsilon_s)+U(n_q-\tfrac12)(n_f-\tfrac12),
 \qquad U=2\Delta.\label{eq:sharpH}
\end{align}
The four displayed unit hoppings set the energy unit. The coupling $v$, offset $\Delta$, and interaction $U$ stay fixed as $\epsilon_s\downarrow0$. Switching off $U$ retains the bare $q$ onsite energy $\epsilon_s+\Delta$.

The $n_f=0$ sector has $q$ energy $\epsilon_s$ and the $n_f=1$ sector has $q$ energy $\epsilon_s+2\Delta$. At infinite temperature,
\begin{equation}
 G_U=\tfrac12G_{\rm res}+\tfrac12G_{\rm off}.
 \label{eq:mixture}
\end{equation}
In the resonant sector, exchange of $(a,b,c)$ and $(q,r,d)$ reduces the one-body problem to
\begin{equation}
 h_\pm=\begin{pmatrix}\epsilon_s&1&0\\1&\pm v&1\\0&1&0\end{pmatrix}.
 \label{eq:parityblocks}
\end{equation}
Its low-energy roots solve $(e-\epsilon_s)[e(e\mp v)-1]-e=0$. Expansion gives
\begin{equation}
 e_\pm=\epsilon_s/2\pm v\epsilon_s^2/8-\epsilon_s^3/16+O(\epsilon_s^4),\qquad
 w_\pm=1/4\pm v\epsilon_s/8+O(\epsilon_s^2).
 \label{eq:split}
\end{equation}
The physical $a$ residue is $w_\pm=\tfrac12\partial_{\epsilon_s} e_\pm$. Both low-energy combinations have connector amplitudes of order $\epsilon_s$, so their fixed bridge produces an avoided crossing of order $\epsilon_s^2$.

For a nonzero fixed bath offset $\delta_B$, elimination of the intermediate orbitals gives
\begin{equation}
 G_{\delta_B}(z)^{-1}=z-\epsilon_s-
 \left[z-\frac1z-\frac{v^2}{z-1/z-1/(z-\epsilon_s-\delta_B)}\right]^{-1}
 =2z-\epsilon_s+(1+v^2)z^3+O(z^4),
 \label{eq:offGF}
\end{equation}
where the expansion is used with $\epsilon_s=O(z)$. Both $\delta_B=\Delta$ (the free reference) and $\delta_B=2\Delta$ (the off-resonant interacting sector) therefore have
\begin{equation}
 e_{\delta_B}=\epsilon_s/2-(1+v^2)\epsilon_s^3/16+O(\epsilon_s^4),\qquad
 Z_{\delta_B}=1/2+O(\epsilon_s^2).
 \label{eq:offpole}
\end{equation}
Let $e_0=e_\Delta$ be the exact free pole. At $\omega=e_0+x\epsilon_s^2$ and $\eta=\gamma\epsilon_s^2$, write $d_v=v/8$. The limiting source spectra are
\begin{align}
 \lim_{\epsilon_s\downarrow0}\epsilon_s^2A_0
 &=\frac1{2\pi}\frac{\gamma}{x^2+\gamma^2},\label{eq:limitfree}\\
 \lim_{\epsilon_s\downarrow0}\epsilon_s^2A_U
 &=\frac1{4\pi}\frac{\gamma}{x^2+\gamma^2}
 +\frac1{8\pi}\left[
 \frac{\gamma}{(x-d_v)^2+\gamma^2}
 +\frac{\gamma}{(x+d_v)^2+\gamma^2}\right].\label{eq:limitint}
\end{align}
The total leading source weight is still $1/2$. Half remains in the center and half splits into side peaks. In particular,
\begin{equation}
 \lim_{\epsilon_s\downarrow0}\frac{A_U(e_0;\gamma\epsilon_s^2)}{A_0(e_0;\gamma\epsilon_s^2)}
 =\frac12+\frac12\frac{\gamma^2}{\gamma^2+(v/8)^2}.
 \label{eq:ratio}
\end{equation}
It is useful to express the resolution by $X=8\eta/(|v|\epsilon_s^2)$. The interaction brings one conserved sector into resonance. A matched quadratic control isolates that sector: set $U=0$ and change the $q$ onsite energy from $\epsilon_s+\Delta$ to $\epsilon_s$, leaving every hopping fixed. Its spectrum is $G_{\rm res}$ in Eq.~\eqref{eq:mixture}. Relative to the original free pole and spectrum, the two limits are
\begin{equation}
 \frac{A_U}{A_0}\longrightarrow\frac12+\frac{X^2}{2(1+X^2)},\qquad
 \frac{A_{\rm res}}{A_0}\longrightarrow\frac{X^2}{1+X^2}.
 \label{eq:crossover}
\end{equation}
At $X=1$ they are $3/4$ and $1/2$. Exact sector-resolvent values were evaluated for $\epsilon_s=0.05,0.02,0.008$. At $\epsilon_s=0.008$, the ratios exceed their respective limits by only $7.1\times10^{-6}$ and $1.4\times10^{-5}$. The quadratic control explains the avoided crossing; the interaction supplies its stationary mixture with the detuned sector. All displayed ratios retain the original free reference.

In the wider but still resolving window $\epsilon_s^2\ll\eta\ll\epsilon_s$, the ratio tends to one. Its leading deficit is
\begin{equation}
 1-\frac{A_U}{A_0}\sim\frac{v^2\epsilon_s^4}{128\eta^2},\qquad
 \R\sim\frac{v^2\epsilon_s^4}{16\eta^2}.
 \label{eq:deficit}
\end{equation}
Thus this thermal resonance lies a factor eight below the leading uniform bound in that window. At finite $\epsilon_s$, subleading residue changes can reverse the sign of the small total deviation. The signed scaled deficit below displays its approach to the positive asymptotic coefficient.

This resonant family realizes the resolution boundary through a coherent splitting of discrete poles: an allowed interacting environment produces order-one spectral changes at $\eta=O(E^2)$, with $E\sim\epsilon_s/2$. Only the source and $q$ onsite energies vary with $\epsilon_s$; all hoppings and interaction strengths stay fixed. It therefore establishes the squared-energy exponent for a vanishing-error guarantee uniform over this class of environments. The companion develops the resulting thermodynamic spectral laws for a fixed disordered Hamiltonian.

For $v=\Delta=1$, $U=2$, direct evaluation of the two conditional six-orbital resolvents gives the following comparison. The second column shows the deviation from $3/4$ at $\eta=\epsilon_s^2/8$, in parts per million. In the last two columns $\eta=\epsilon_s^{3/2}$ and $S=(1-A_U/A_0)\eta^2/\epsilon_s^4$:
\begin{center}
\begin{tabular}{cccc}
\toprule
$\epsilon_s$&$10^6(A_U/A_0-3/4)$&$S$&$\R\eta^2/\epsilon_s^4$\\
\midrule
0.050&290&$-0.001627$&0.09022\\
0.010&11.1&0.005931&0.06764\\
0.005&2.75&0.006873&0.06505\\
0.002&---&0.007437&0.06352\\
0.001&---&0.007625&0.06301\\
0.0005&---&0.007719&0.06276\\
0.0002&---&0.007775&0.06261\\
$\epsilon_s\to0$&$0$&$1/128$&$1/16$\\
\bottomrule
\end{tabular}
\end{center}
Every entry is evaluated at the exact pole of the predetermined free Hamiltonian. The two scaled quantities approach finite coefficients with ratio $1/8$.

\subsection{Propagation and transition factors in one bound}
\label{sec:unified}

An environmental process changes the source response only after propagating through the contact and back. A free propagation zero and a restriction on actual target transitions can each suppress this return process. We now combine these two factors. The integer $n$ counts the frequency factor imposed on each target transition: none for an unconstrained target ($n=0$), and one for a resonant side orbital ($n=1$).

Let $C$ be the fixed number-conserving quadratic block, $P_a$ select the measured source, and $P_q$ select its free contact orbitals. The bridge is $q^\dagger Vr+\hc$, and $F_\alpha=G_{rr}^{(\alpha)}$ is evaluated in the full coupled system. For an unconstrained target, take $n=0$, $X_0=V$, and $M_\alpha=F_\alpha$. For $n=1$, canonical side orbitals $d$ have only the terms $\Omega d^\dagger d+r^\dagger Sd+\hc$, and the bridge factors as $V=X_1S^\dagger$. The matrix $S$ may be rectangular. Writing $\xi=z-\Omega$ and $Q_\alpha=G_{dd}^{(\alpha)}$, its exact equation of motion gives
\begin{equation}
 M_\alpha=S^\dagger F_\alpha S=\xi^2Q_\alpha-\xi I,
 \qquad M_1-M_0=\xi^2(Q_1-Q_0)\quad(n=1).
 \label{eq:unifiedtarget}
\end{equation}
The common term cancels from the difference. For a single environment, the canonical ball in Eq.~\eqref{eq:ball2} maps to a ball of radius $|\xi|^2/(2\eta)$ centered at $-\ii|\xi|^2I/(2\eta)$. Thus both target cases obey $\|M_1-M_0\|,\|M_\alpha\|\le|\xi|^{2n}/\eta$. Only the free side-mode equation is shifted by $\Omega$; bath pairing remains allowed.

Suppose the free contact equations give
\begin{equation}
 L_n(z):=X_n^\dagger P_qg_C(z)P_a^\dagger
 =Z(z)P_\ell g_C(z)P_a^\dagger,\qquad \|P_\ell\|\le1,
 \label{eq:unifiedfactor}
\end{equation}
where $P_\ell$ selects specified free coordinates. The contact equations determine $Z$ from the fixed local matrices. The exact source correction is $L_n(\bar z)^\dagger M_\alpha(z)L_n(z)$. Applying Eq.~\eqref{eq:ward} to its two external legs yields
\begin{align}
 |u^\dagger(G_{aa}^{(1)}-G_{aa}^{(0)})u|
 &\le\delta_n(z)[-\ImH(u^\dagger g_{C,aa}u)],\label{eq:unifiedproof}\\
 \delta_n(z)&=\frac{|\xi|^{2n}\|Z(z)\|\|Z(\bar z)\|}{\eta^2}.
 \label{eq:unifieddelta}
\end{align}
The single-environment estimate gives $(1-\delta_n)\A_C\preceq\A_\alpha$. Therefore Eqs.~\eqref{eq:mainbound} and \eqref{eq:positivebound} hold with $\delta=\delta_n<1$, relative to the complete reference $\A_0$. If both $Z$ norms are bounded by $c_m|\xi|^m$, the combined order is $p=m+n$: the sufficient resolving window is $c_mE^p\ll\eta\ll E$ for $p>1$, where $E=|\omega-\Omega|$ and $c_m$ has units of energy$^{1-p}$.

For the two-side contact, let $V=S_1MS_2^\dagger$, choose $X_1=S_1M$, and use the first side-mode row $S_1^\dagger P_bg_CP_a^\dagger=\xi P_cg_CP_a^\dagger$. Then $Z=\xi M^\dagger$, $P_\ell=P_c$, and
\begin{equation}
 (m,n)=(1,1),\qquad \delta_1=\|M\|^2|\xi|^4/\eta^2.
 \label{eq:unifiedtwoleaves}
\end{equation}
For invertible side contacts this is exactly Eq.~\eqref{eq:kappa}. The companion develops contacts that supply both powers through free propagation, with $(m,n)=(2,0)$, as well as multifrequency resonator designs. The present contact supplies the two powers through distinct transition and propagation constraints.

\section{How virtual polarization transfers collective motion to the source}
\label{sec:collective}

A protected particle spectrum would be uninteresting as a collective probe if the source were simply decoupled. The missing ingredient is a response of the electrons already occupying the device. A slow environmental spin field changes their orbital composition through virtual mixing with other bands. We calculate how much spin this transfers to the source and show why the resulting orbital correlations follow the environmental motion.

The free-band calculation fixes the leading transfer coefficient before any coupled-model response is measured. The next two sections compare this prediction with a quantum device and a two-dimensional lattice. We return to the feedback of the metal on critical spin dynamics in Sec.~\ref{sec:collective-feedback}.

\subsection{A metal coupled to quantum Ising spins}
\label{sec:collective-model}

Repeat the contact of Eq.~\eqref{eq:H} at every source site and introduce two fermion spin species. In the orbital order $(a,b,c,r,d)$, the free Bloch matrix for either species is
\begin{equation}
 h_0(\xi_{\bm k})=
 \begin{pmatrix}
 \xi_{\bm k}&g&0&0&0\\
 g&0&s_1&v&0\\
 0&s_1&0&0&0\\
 0&v&0&0&s_2\\
 0&0&0&s_2&0
 \end{pmatrix}.
 \label{eq:collective-h}
\end{equation}
The source dispersion $\xi_{\bm k}$ is measured from the common side-orbital resonance. All contact parameters are real and nonzero. Quantum spins $\tau_i$ couple to the target spin density:
\begin{align}
 H_\tau&=-J_\tau\sum_{\langle ij\rangle}\tau_i^z\tau_j^z
              -h_\tau\sum_i\tau_i^x,\nonumber\\
 H_{\rm int}&=\lambda\sum_i\tau_i^z M_{r_i},
 &M_{\alpha_i}&=n_{\alpha_i\uparrow}-n_{\alpha_i\downarrow}.
 \label{eq:collective-spin}
\end{align}
The transformation $\tau^z\mapsto-\tau^z$, accompanied by exchange of the two fermion species, is an exact Ising symmetry. The transverse field makes the target potentials fluctuate through the quantum dynamics of the Ising spins.

Equation~\eqref{eq:collective-spin} respects the support condition of the spectral theorem: the interaction acts only on $r$ and the environment. An entirely fermionic representation uses two additional bath orbitals $f_{i+},f_{i-}$ in the invariant sector $n_{i+}+n_{i-}=1$, with $\tau_i^z=n_{i+}-n_{i-}$ and $\tau_i^x=f_{i+}^\dagger f_{i-}+\hc$. Adding or removing a source fermion preserves that sector. The side-orbital equations and canonical spectral sum rule used in Sec.~\ref{sec:proof} therefore apply to the quantum spin environment without change.

For a concrete metallic reference, take
\begin{equation}
 \xi_{\bm k}=-2t_s(\cos k_x+\cos k_y)+\epsilon_s,
 \qquad t_s=\epsilon_s=0.05J,
 \qquad g=s_1=s_2=v=J.
 \label{eq:collective-lattice}
\end{equation}
Its Fermi contour is $\cos k_x+\cos k_y=1/2$, away from the van Hove energy. The Fermi velocity is nonzero and the density of states is finite. At $\xi=0$, the nearest other band has energy magnitude $J\sqrt{2-\sqrt2}$. Since $|\xi_{\bm k}|\le0.25J$, the four other bands remain separated from zero by at least
\begin{equation}
 \Delta_h\ge J\sqrt{2-\sqrt2}-0.25J>0.515J.
 \label{eq:collective-gap}
\end{equation}
This analytical lower bound supplies the band separation used below. The Ising energy scale and $|\lambda|$ can be chosen below it.

\subsection{Why low-energy electrons couple weakly to the order parameter}
\label{sec:collective-node}

The metal must remain connected while its low-energy electrons avoid the interacting target. The five orbital equations show how both requirements are met: a dispersive band crosses the Fermi level, but its target component vanishes there. We extract that component explicitly.
\begin{equation}
 Z=\frac{s_1^2}{s_1^2+g^2},\quad
 R(E)=E^2-s_2^2,\quad
 B(E)=(E^2-s_1^2)(E^2-s_2^2)-v^2E^2.
\end{equation}
The inverse dispersion and an eigenvector with its $a$ component set to one are
\begin{align}
 \xi=D(E)&=E-\frac{g^2E R(E)}{B(E)},\nonumber\\
 u(E)&\propto\left(1,\frac{gE R(E)}{B(E)},
       \frac{gs_1R(E)}{B(E)},\frac{gvE^2}{B(E)},
       \frac{gvs_2E}{B(E)}\right).
 \label{eq:collective-band}
\end{align}
The central band crosses the Fermi energy and is the low band used below. After normalizing its eigenvector, its energy is $e_{\bm k}=Z\xi_{\bm k}+O(\xi_{\bm k}^3)$ and its target weight obeys
\begin{equation}
 |u_r(e_{\bm k})|^2=C_r e_{\bm k}^4+O(e_{\bm k}^6),
 \qquad C_r=\frac{Zg^2v^2}{s_1^4s_2^4}.
 \label{eq:collective-node}
\end{equation}
Thus the target \emph{amplitude} vanishes quadratically everywhere on the Fermi contour. The total density of low-energy electronic states remains finite. It is the component seen by the Ising spins that vanishes.

Let $\psi_k$ annihilate a fermion in this low band, with $k$ including momentum and imaginary frequency, and let $\phi$ denote the Ising field. Its direct vertex is
\begin{equation}
 \lambda\phi_q F_{\bm k+\bm q}^*F_{\bm k}\,
 \bar\psi_{k+q}\psi_k,
 \qquad F_{\bm k}=u_r(e_{\bm k})=O(e_{\bm k}^2).
 \label{eq:collective-vertex}
\end{equation}
Each incoming or outgoing low-band fermion brings its own small target amplitude. This suppresses the decay of a spin fluctuation into a low-energy particle--hole pair.

\subsection{Virtual transitions give the source a critical spin response}
\label{sec:collective-matching}

A slowly varying target potential mixes occupied eigenstates with other bands and redistributes their density among orbitals. This virtual mixing changes orbital polarization while keeping the particles in their occupied states. Opposite target potentials for the two species induce an orbital spin. The derivative of the free density with respect to the potential gives the transferred spin per unit Ising field.

To determine the amount of spin transferred to each orbital, first consider one free fermion species. Add a uniform target onsite energy $m$ and define the zero-temperature, fixed-chemical-potential response
\begin{equation}
 \kappa_\alpha=
 \left.\frac{\partial\langle n_\alpha\rangle}{\partial m}\right|_{m=0}.
 \label{eq:collective-kappadef}
\end{equation}
The effective low-energy representation of a physical orbital spin density is then
\begin{equation}
 M_\alpha=b_\alpha\phi+
 \text{regular fermion bilinears and less singular operators},
 \qquad b_\alpha=2\lambda\kappa_\alpha+O(\lambda^3).
 \label{eq:collective-matching}
\end{equation}
Here $\phi$ is normalized as the long-wavelength microscopic Ising field. The two fermion species feel opposite target potentials, so their induced density changes add in the spin difference, giving the factor two. Within the low band, the response connecting the target field to an orbital density retains four low-energy powers from Eq.~\eqref{eq:collective-vertex} and has no constant term. Virtual transitions between bands supply the constant $\kappa_\alpha$ and hence the leading overlap with the Ising field.

For an isolated cell at $\xi=0$, set $\Omega_c=\sqrt{s_1^2+g^2}$. The mixed resolvent gives an explicit nonzero source coefficient:
\begin{align}
 \kappa_a&=-\frac{1}{\pi}\int_0^\infty
 \frac{g^2v^2\omega^2\,\dd\omega}
 {[\omega^4+(\Omega_c^2+v^2+s_2^2)\omega^2+
 \Omega_c^2s_2^2]^2}\nonumber\\
 &=-\frac{g^2v^2}
 {4\Omega_c|s_2|[(\Omega_c+|s_2|)^2+v^2]^{3/2}}<0.
 \label{eq:collective-kappaexact}
\end{align}
The dark-mode occupation is fixed and contributes no target-field response. Continuity gives a nonzero coefficient for a finite interval of source bandwidths. For the metallic lattice in Eq.~\eqref{eq:collective-lattice}, the band integral gives
\begin{equation}
 \begin{aligned}
 (\kappa_a,\kappa_b,\kappa_c,\kappa_r,\kappa_d)J
 \simeq{}&(-0.009530,\;0.01405,\;-0.01005,\\
 &\hspace{12mm}-0.2111,\;0.2166).
 \end{aligned}
 \label{eq:collective-kappavalues}
\end{equation}
The $240\times240$ momentum integral fixes the leading response coefficient before the interacting calculation. Between $60\times60$ and $240\times240$ grids, $\kappa_a$ changes by less than $3\times10^{-12}/J$; centered field differences agree within $2.4\times10^{-10}/J$.

Because the source spin contains a component proportional to the Ising field, its long-distance correlations inherit the Ising singularity. At the Ising transition, Eq.~\eqref{eq:collective-matching} gives
\begin{equation}
 \chi_{M_aM_a}^{\rm sing}(\bm q,\ii\Omega)
 =b_a^2\chi_{\phi\phi}(\bm q,\ii\Omega)
 \propto b_a^2(\Omega^2+c_{\rm I}^2q^2)^{-1+\eta_{\rm I}/2}.
 \label{eq:collective-sourcechi}
\end{equation}
Here $c_{\rm I}$ is the critical spin velocity. In an ordered state, $\langle M_a\rangle=b_a\langle\phi\rangle+\cdots$. The same physical source that preserves its one-particle spectrum therefore develops a singular collective response.

The two measurements now have explicit predictions. Spin response contains the environmental susceptibility with coefficient $b_a^2$, whereas the addition--removal spectrum obeys Eq.~\eqref{eq:mainbound}. This is the promised coexistence on a physical orbital, with both the spectral error and the transferred spin amplitude specified.

\begin{figure}[htbp]
\centering
\includegraphics[width=.56\textwidth]{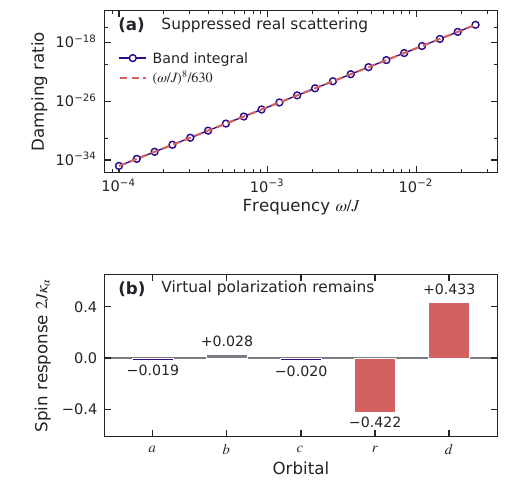}
\caption{Free-band inputs to the collective mechanism. (a) The intraband damping ratio approaches $(\omega/J)^8/630$ for the matched two-side and no-side metals. (b) Virtual-polarization coefficients $2J\kappa_\alpha$ in the periodic metallic reference. These are band calculations; the independent coupled-model responses are given in Sec.~\ref{sec:source-full-quantum-test}.}
\label{fig:collective-free-inputs}
\end{figure}

\subsection{Why the singular spin polarization is compensated within a cell}
\label{sec:collective-compensation}

The orbital polarization has a simple spatial pattern: spin gained on one orbital is lost on others in the same cell. This follows because a target field changes the composition of occupied states without moving the Fermi contour. Summing the density derivatives over all orbitals removes the redistribution terms and leaves only a possible change of occupation at the Fermi level:
\begin{equation}
 \sum_\alpha\kappa_\alpha
 =-\sum_\nu\int_{\bm k}\delta(E_{\nu\bm k})
 |u_{\nu r}(\bm k)|^2=0.
 \label{eq:collective-kappasum}
\end{equation}
The last equality uses the target node at the Fermi contour. Wavefunction redistributions between occupied orbitals cancel by normalization. More generally, for a uniform finite target potential $m$,
\begin{equation}
 \det h_m(\xi)=\xi s_1^2s_2^2,
 \label{eq:collective-staticdet}
\end{equation}
so the free Fermi contour remains at $\xi=0$. The compensation displayed numerically in Eq.~\eqref{eq:collective-kappavalues} therefore follows from the exact free-band identity in Eq.~\eqref{eq:collective-kappasum}.

Conduction-spin conservation gives a complementary dynamic constraint. Define $Q=\sum_{i,\alpha}M_{i\alpha}$, excluding the separate Ising spins. Since $[Q,H]=[Q,\tau_j^z]=0$,
\begin{equation}
 \chi^R_{Q,\tau_j^z}(t)
 =-\ii\theta(t)\langle[Q,\tau_j^z(t)]\rangle=0.
 \label{eq:collective-ward}
\end{equation}
The mixed Matsubara response vanishes at nonzero frequency. The zero-frequency thermodynamic covariance is a different limit.

Within the local low-energy matching used above, a term proportional to the Ising order field in the total conduction density would produce a dynamic response forbidden by the Ward identity. The metallic nonlocal mixed response cannot supply a cancelling constant: it retains four energy powers at its target vertex, as established by the external-factor argument in Sec.~\ref{sec:collective-factorization}. Consequently the leading critical overlap satisfies $\sum_\alpha b_\alpha=0$ in this low-energy theory. The singular orbital response matrix is
\begin{equation}
 \chi_{\alpha\beta}^{\rm sing}
 =b_\alpha b_\beta\chi_{\phi\phi},
 \qquad \chi^{\rm sing}\bm 1=0.
 \label{eq:collective-rankone}
\end{equation}
It has rank one and represents a compensated polarization pattern within the cell. Regular Pauli response and distinct static thermodynamic limits remain separate from this leading Ising singularity.


\section{Particle and spin responses in the same quantum device}
\label{sec:quantum-device}

The most direct demonstration uses one state and two probes. We solve a device containing both mobile electrons and quantum Ising spins, identify the environmental excitation in the source-spin response, and calculate the particle spectrum of that same source. Increasing the coupling changes the collective excitation substantially. The key comparison is then its change beside the particle-spectrum change at one common resolution.

\subsection{Both responses in the same coupled quantum Hamiltonian}
\label{sec:source-full-quantum-test}

Two cells are enough to give the Ising spins a collective excitation and the electrons a connected multiband structure. All four intracell contacts have strength $J$, the source orbitals have zero onsite energy and hopping $t_s=0.05J$, and $J_\tau=h_\tau=0.02J$. We use the lowest state in the sector $N_\uparrow=N_\downarrow=5$, retaining every fermion band and both quantum spins. The sector contains
\begin{equation}
 \binom{10}{5}^2\,2^2=254016.
\end{equation}

To identify the environmental excitation in the source response, let $|0\rangle$ be the lowest state in this fixed-number sector and $|p\rangle$ the low-energy excited state carrying the largest weight of $\Phi=(\tau_1^z+\tau_2^z)/\sqrt2$. For $M_a=(M_{a_1}+M_{a_2})/\sqrt2$, define
\begin{equation}
 b_{\rm eff}=\frac{\langle0|M_a|p\rangle}
 {\langle0|\Phi|p\rangle},\qquad
 W_\Phi=|\langle0|\Phi|p\rangle|^2.
 \label{eq:source-measured-b}
\end{equation}
The source pole weight is $W_a=b_{\rm eff}^2W_\Phi$. These three quantities are obtained from the full interacting eigenvectors. They identify the same excitation in the source and the Ising spins, so the particle and collective comparisons refer to one physical state.

\subsection{A strongly shifted collective mode beside a stable particle peak}
\label{sec:strong-coupling-response}

Figure~2(a,b) of the Letter compares the complete particle spectrum and complete source-spin absorption with the common half-width $\eta=\gamma=0.003125J$. We first follow the collective excitation as the interaction grows, and then calculate the two spectra of each state. The isolated two-spin gap is
\begin{equation}
 \Delta_\tau^{(0)}
 =\sqrt{J_\tau^2+4h_\tau^2}-J_\tau
 \simeq0.02472J.
 \label{eq:strong-isolated-gap}
\end{equation}
This is a finite-system tunneling splitting. We follow the corresponding interacting pole through its largest uniform Ising matrix element and calculate the source-spin weight $W_a=|\langle0|M_a|p\rangle|^2$ at that same pole. The full quantum calculation gives
\begin{center}
\begin{tabular}{rrrrr}
\toprule
$\lambda/J$ & $\Delta_p/J$ & Gap reduction & $W_a$ & $b_{\rm eff}$\\
\midrule
0.1 & 0.02465 & $0.286\%$ & $5.572\times10^{-6}$ & $-0.001962$\\
0.5 & 0.02304 & $6.810\%$ & $1.315\times10^{-4}$ & $-0.009474$\\
1 & 0.01888 & $23.61\%$ & $4.448\times10^{-4}$ & $-0.01714$\\
2 & 0.009799 & $60.4\%$ & $1.045\times10^{-3}$ & $-0.02500$\\
\bottomrule
\end{tabular}
\end{center}
At $\lambda=2J$, the interaction changes the collective energy by $60\%$ and increases its source-spin absorption weight by a factor $187$ relative to $\lambda=0.1J$. The measured source overlap differs from the linear weak-coupling prediction by $36\%$. The table values are extracted directly from the full interacting eigenvectors.

The collective pole establishes that the source spin participates in the environmental motion. To determine what a particle probe sees in this same state, we calculate the complete local $a_1\uparrow$ addition--removal resolvent,
\begin{align}
 G_a(z)={}&\langle v_+|[z+E_0-H_{6,5}]^{-1}|v_+\rangle\nonumber\\
 &+\langle v_-|[z-E_0+H_{4,5}]^{-1}|v_-\rangle,
 \label{eq:source-adjacent-resolvent}\\
 |v_+\rangle={}&a_{1\uparrow}^\dagger|0\rangle,\qquad
 |v_-\rangle=a_{1\uparrow}|0\rangle.\nonumber
\end{align}
Each adjacent sector has 211680 states. Lanczos tridiagonalization starts from the actual addition and removal vectors, approximating the complete resolvent without discarding high-energy states or Lorentzian tails. The reference is the complete connected free Hamiltonian at the same one-body parameters.

At $\lambda=2J$, the collective gap has fallen from $0.02472J$ to $0.009799J$, a reduction of $60.4\%$. At the free particle peak $E_0\simeq0.02498J$, the full particle intensity changes by $-3.594$ parts per million at this same resolution. Across the sampled bright interval $[0.02J,0.03J]$, its maximum fractional change is $9.346$ ppm. The reference peak signal is $\pi\eta\mathcal A_0(E_0)=0.2505$, and the uniform theorem bound is below $9.27\%$ throughout the interval. The numerical spectrum gives the much smaller change realized by this device; the theorem supplies the guarantee over all allowed environments.

The collective pole lies $3.14$ half-widths from zero, so the shared resolution resolves its softening. The next calculation includes the rest of the absorption spectrum to determine how this pole appears in the measured response.

\subsection{Complete collective absorption at the same resolution}
\label{sec:full-collective-absorption}

The absorbed power includes every transition reached by the source-spin probe. We evaluate their sum in the same ground state and at the same resolution as the particle spectrum. For $M_a=2^{-1/2}\sum_{j=1}^2(n_{a,j,\uparrow}-n_{a,j,\downarrow})$, write $\Delta_n=E_n-E_0$ and $W_n=|\langle n|M_a|0\rangle|^2$. The positive-absorption convention gives
\begin{equation}
 S_a(\omega;\gamma)\equiv\frac{\chi''_{aa}(\omega;\gamma)}{\pi}
 =\sum_{n>0}W_n\left[L_\gamma(\omega-\Delta_n)
                         -L_\gamma(\omega+\Delta_n)\right],
 \label{eq:full-collective-positive}
\end{equation}
where $L_\gamma(x)=\gamma/[\pi(x^2+\gamma^2)]$.
Both frequency partners contribute even at positive probe frequency.
To retain all excited states, we start Lanczos iteration from
$|s\rangle=(M_a-\langle M_a\rangle)|0\rangle$ and evaluate
$F(z)=\langle s|[z-(H-E_0)]^{-1}|s\rangle$.
Then $S_a=-\operatorname{Im}[F(\omega+i\gamma)+F(-\omega-i\gamma)]/\pi$.
The seed retains its overlap with the full excited-state space.

At $\gamma=\eta=0.003125J$, the strong-coupling case $\lambda=2J$
has a full absorption maximum $S_a=0.1038/J$ at
$\omega=0.009811J$.
At the selected pole frequency $\Delta_p=0.009799J$, the rest of the
many-body spectrum contributes $0.01236\%$ of the total response.
It contributes $0.1029\%$ of the absorption integrated over the fixed band
$\mathcal B=[0.005,0.035]J$.
At $\omega=\Delta_p$, the negative-frequency partner reduces the pole
contribution by $2.480\%$ compared with a positive Lorentzian alone.
The complete collective peak thus remains visible at the same resolution
at which the complete particle spectrum obeys the stated relative bound.

\begin{table}[ht]
\centering
\caption{Complete source-spin absorption at $\gamma=0.003125J$.
The background comprises all transitions except the selected pole and its
negative-frequency partner. The peak fraction is evaluated at $\Delta_p$;
the band fraction uses the full integral over $\mathcal B$.}
\begin{tabular}{ccrr}
\hline
$\lambda/J$ & $\omega_{\max}/J$ & Background at pole & Background in band\\
\hline
0.1 & 0.02466 & $4.534\%$ & $12.26\%$\\
0.5 & 0.02304 & $0.1908\%$ & $0.5933\%$\\
1 & 0.01889 & $0.04842\%$ & $0.1840\%$\\
2 & 0.009811 & $0.01236\%$ & $0.1029\%$\\
\hline
\end{tabular}
\label{tab:complete-collective-background}
\end{table}

The weaker signal at $\lambda=0.1J$ has a larger relative background,
as Table~\ref{tab:complete-collective-background} shows.
To understand the low-frequency contribution of higher-energy transitions,
we combine their positive- and negative-frequency partners. Their tails
partially cancel:
\begin{equation}
 L_\gamma(\omega-\Delta)-L_\gamma(\omega+\Delta)
 =\frac{4\gamma\omega\Delta}
 {\pi[(\omega-\Delta)^2+\gamma^2][(\omega+\Delta)^2+\gamma^2]}.
 \label{eq:collective-tail-pair}
\end{equation}
Thus high-energy transitions contribute as $4\gamma\omega/\pi\Delta^3$.
More generally, a group with total weight $W_{>}$ and
$\Delta\ge\Lambda>\omega$ contributes at most
$4\gamma\omega\Lambda W_{>}/[\pi(\Lambda^2-\omega^2)^2]$.
The reported spectra evaluate those contributions directly.

\subsection{One-side controls in the same quantum environment}
\label{sec:strong-one-side-control}

The benefit of the second side orbital is best assessed against a device that also transmits a collective signal. We therefore compare with the chain $a$--$r$--$d$, keeping both bonds equal to $J$. It retains the same source residue $Z=1/2$, leading low-band velocity, source hopping, source onsite energy, and quantum Ising Hamiltonian. Each species remains mobile and couples through $\lambda\tau_i^zM_{r_i}$. The half-filled sector now has $N_\uparrow=N_\downarrow=3$ and 1600 states. Each contact is compared with its own connected free reference.

Figure~2(c) uses $\eta=0.005J$ for the contact comparison. At this width, the spectral comparison at each bright free resonance and the same interaction strength is
\begin{center}
\begin{tabular}{rrrr}
\toprule
$\lambda/J$ & Two-side error (ppm) & One-side error (ppm) & $W_a^{(1)}/W_a^{(2)}$\\
\midrule
0.5 & 0.2040 & 154.6 & 72.60\\
1 & 0.8227 & 679.3 & 79.81\\
2 & 3.373 & 4036 & 106.6\\
\bottomrule
\end{tabular}
\end{center}
At $\lambda=2J$, the one-side pole lies at $0.01339J$ and has source weight $0.1113$. The two-side contact reduces the bright-peak spectral distortion by a factor $1197$, while also reducing the source collective weight. The reduction in collective weight changes the available readout signal. On the entire common bright window $[0.02J,0.03J]$, the maximum one-side error is $0.8384\%$, compared with $6.625$ ppm for two sides.

At equal coupling, the two-side contact has both a smaller particle-spectrum distortion and a smaller collective signal. To measure the gain in spectral protection at the same collective signal, we match the actual source-spin absorption weight by varying only the one-side value of $\lambda$. The bath Hamiltonian and all one-body parameters remain fixed. This gives
\begin{center}
\begin{tabular}{rrrr}
\toprule
$\lambda_{\rm two}/J$ & $\lambda_{\rm one}/J$ & Common $W_a$ & Error ratio (one/two)\\
\midrule
1 & 0.1066 & $4.448\times10^{-4}$ & 8.305\\
2 & 0.1635 & $1.045\times10^{-3}$ & 4.774\\
\bottomrule
\end{tabular}
\end{center}
At matched integrated collective weight, the two-side contact reduces the particle-spectrum distortion by a factor $4.8$--$8.3$. Using the stored full-precision matching couplings, the weights agree to a relative error below $2.1\times10^{-11}$, and the free peak intensities are close: $\pi\eta\mathcal A_0=0.2520$ for two sides and $0.2522$ for one side. The matched poles retain their own frequencies: $0.01888J$ versus $0.02467J$ for the first pair, and $0.009799J$ versus $0.02460J$ for the second. Their integrated weights define the matching used in the Letter. The companion gives the corresponding static and finite-band responses. The uniform distinction between one and two zeros is established by the theorem and the counterexamples in Sec.~\ref{sec:geometry-hierarchy-round9}.

\subsection{Connecting the quantum response to virtual polarization}

The strong-coupling comparison establishes the separation directly. At weaker coupling, the free-band calculation also predicts how much spin reaches the source. This gives a quantitative link between the mechanism of Sec.~\ref{sec:collective} and the quantum eigenstates.

The free two-cell virtual-polarization coefficient is $\kappa_a^{\rm 2cell}\simeq-0.009816/J$; the last column below compares the measured overlap with its weak-coupling prediction.

\begin{center}
\begin{tabular}{rrrr}
\toprule
$\lambda/J$ & $(E_p-E_0)/J$ & $b_{\rm eff}$ & $b_{\rm eff}/(2\lambda\kappa_a^{\rm 2cell})-1$ (\%)\\
\midrule
0.025 & 0.02472 & $-0.0004911$ & $+0.062$\\
0.05 & 0.02470 & $-0.0009820$ & $+0.034$\\
0.10 & 0.02465 & $-0.001962$ & $-0.076$\\
0.20 & 0.02444 & $-0.003906$ & $-0.516$\\
\bottomrule
\end{tabular}
\end{center}

At $\lambda=0$ the source residue at the Ising pole is below $2\times10^{-32}$. It becomes $5.572\times10^{-6}$ at $\lambda=0.1J$ and $2.213\times10^{-5}$ at $0.2J$. These finite-frequency pole weights establish low-frequency collective absorption directly in the coupled quantum system. The static matching coefficient predicts its leading weak-coupling overlap. At fixed $\lambda=0.1J$, reducing $h_\tau/J$ from $0.02$ to $0.01$ and $0.005$ moves this pole to $0.008256J$ and $0.002352J$. The corresponding matching ratios are $0.9986$ and $0.9985$: the source follows the softening collective excitation. For comparison, the isolated pair has the exact gap $\sqrt{J_\tau^2+4h_\tau^2}-J_\tau$. Its decrease is a finite-system tunneling splitting. The calculation tests the transfer of this excitation to the source; the extended critical application is developed in Sec.~\ref{sec:collective-feedback}.

The pole also resolves the compensated polarization directly. At $\lambda=0.1J$, its overlap vector in the order $(a,b,c,r,d)$ is
\begin{equation}
 \begin{aligned}
 \bm b_{\rm eff}\simeq{}&(-0.001962,\ 0.002801,\\
 &-0.001986,\ -0.04220,\ 0.04334).
 \end{aligned}
\end{equation}
Before rounding, the sum vanishes within $1.1\times10^{-17}$. The low-frequency source spin moves by redistributing polarization among the orbitals while their total conserved spin remains fixed.

\subsection{Numerical method and precision of the comparison}
\label{sec:device-precision}

The fixed-number construction retains all bands and all quantum-spin states. Particle spectra use adjacent-sector resolvents; collective absorption uses the vector $(M_a-\langle M_a\rangle)|0\rangle$. Both include the full spectral tails. The checks below quantify different parts of this calculation.

\begin{center}
\begin{tabular}{p{0.29\textwidth}p{0.61\textwidth}}
\toprule
Quantity & Numerical check\\
\midrule
Strong-coupling states & Ground-state residuals below $2.6\times10^{-12}J$; retained collective-state residuals below $1.5\times10^{-10}J$.\\
Particle normalization & Addition and removal weights sum to one within $9\times10^{-15}$.\\
Particle resolvent & Spectra from 100--500 Lanczos steps agree at the displayed precision. At zero coupling, the full-sector result agrees with the direct free resolvent to $4.1\times10^{-12}$ in the spectral ratio over $|\omega|\le0.06J$.\\
Complete spin absorption & 100--600 steps agree at displayed precision. Zeroth and first spectral moments agree with direct expectations to $4\times10^{-16}$ and $1.3\times10^{-15}J$.\\
One-side control & Dense diagonalization of the 1600-state matched control agrees with its continued fraction to $2.9\times10^{-13}/J$ at the common resolution.\\
\bottomrule
\end{tabular}
\end{center}

These convergence comparisons are supplemented by a direct bound on the error in the ppm-scale particle intensity. The full tridiagonal coefficients, transition matrix elements, and model parameters are supplied with the data.

\paragraph{Direct residual control of the particle spectrum.}
\label{sec:particle-precision}
The particle change is only a few parts per million, so we compare its numerical error directly with that signal. A residual in the original Hilbert space bounds the complete resolvent error, including all directions outside the Krylov basis. For the computed ground state $|g\rangle$ and energy $e$, define $v_+=a^\dagger|g\rangle$, $v_-=a|g\rangle$, $K_+=H_{N+1}-e$, and $K_-=e-H_{N-1}$. We reconstruct $x_\pm$ from 200 fully reorthogonalized Krylov vectors and evaluate
\begin{equation}
 r_\sigma=v_\sigma-(z-K_\sigma)x_\sigma,
 \qquad
 \left|v_\sigma^\dagger(z-K_\sigma)^{-1}v_\sigma
       -v_\sigma^\dagger x_\sigma\right|
 \le\frac{\|v_\sigma\|\,\|r_\sigma\|}{\eta}.
 \label{eq:particle-direct-residual}
\end{equation}
Indeed, the exact solution obeys $x_{\sigma,\mathrm{exact}}-x_\sigma=(z-K_\sigma)^{-1}r_\sigma$. Hermiticity gives $\|(z-K_\sigma)^{-1}\|\le1/\eta$, so the residual controls the error in the full Hilbert space, including every direction omitted from the Krylov basis. Summing its addition and removal contributions bounds
the error in $\mathcal A/\mathcal A_0$ by
$\sum_\sigma\|v_\sigma\|\|r_\sigma\|/(\pi\eta\mathcal A_0)$.

At $\lambda=2J$ and $\eta=0.003125J$, we check the free bright peak,
both ends of $[0.02J,0.03J]$, and the previously sampled maximum at
$0.02793J$. The residuals are below $1.0\times10^{-12}$ for addition
and $6.7\times10^{-14}$ for removal. At the free peak the signed
spectral change is $-3.594$ ppm, with a direct residual bound of
$3.0\times10^{-6}$ ppm. The corresponding bound is below
$5.8\times10^{-6}$ ppm at all four checked frequencies.

We also propagate numerical source-state and reference errors.
The ground-state residual is below $3.0\times10^{-14}J$, and independent
new and stored low-state solves give a gap $0.009799J$.
Let $|g_0\rangle,E_g$ be the exact ground state and energy, and let $\rho=\|(H-e)|g\rangle\|$. With the numerically identified isolation margin $d=0.0097J$, the residual gives $\|g-g_0\|\le\sqrt2\rho/d$ (after phase alignment) and $|e-E_g|\le\rho$. These errors enter in two ways. The state error changes the addition and removal source vectors; the energy error shifts the resolvent argument. The canonical anticommutation relation bounds the combined source-vector norm, while the resolvent identity bounds the energy shift. Their relative spectral contributions are, respectively, the two terms in
\begin{equation}
 \frac{2\sqrt2\rho/(d\eta)+\rho/\eta^2}
      {\pi\mathcal A_0}.
 \label{eq:particle-source-state-error}
\end{equation}
With conservative floating-point allowances and a direct free-reference
residual check, the resulting error budget is below $0.006$ ppm at the
peak and $0.019$ ppm at all four checked frequencies. This last budget
uses the numerically identified isolated ground state; the bound in
Eq.~\eqref{eq:particle-direct-residual} holds for the computed source
vectors without a spectral-gap input. Thus the precision check resolves
the reported $3.6$ ppm intensity change. The interval maximum is the sampled value specified above.



\section{Collective participation in the two-dimensional coupled model}
\label{sec:qmc-scaling}

The two-cell device establishes that the protected source can carry a strongly shifted collective excitation. In a lattice, the next question is spatial: does the source also share spin fluctuations that extend over more cells? We measure the Ising correlations and their overlap with the source in the full coupled Hamiltonian. Following the longest nonuniform wavelength as the lattice grows reveals the growth of the source's collective component.

\subsection{Source participation at a fixed field}
\label{sec:qmc-source-growth}

At $h_\tau=0.061J$, the source component grows nearly fivefold between $L=4$ and $10$, as shown in Fig.~3 of the Letter. Specifically, $J\chi_{\parallel}$ rises from $7.35(19)\times10^{-5}$ to $3.65(5)\times10^{-4}$, while the Ising response $J\chi_{\Phi\Phi}$ rises from $19.53(7)$ to $102.26(23)$. Every size uses $\lambda=0.1J$, $\beta J=50L$, and $\Delta\tau J=0.0625$. We next define the response component and explain why it measures the source's participation in those fluctuations.

The measurement follows the longest nonuniform wavelength of each lattice, so its spatial scale increases with $L$. The source retains a finite overlap with the growing spin fluctuations. Its magnitude follows the virtual-polarization coefficient calculated independently from the free bands in Sec.~\ref{sec:qmc-free-transfer}. The definitions below identify exactly which part of the source response is measured.

\subsection{Which component of the source response follows the spins?}
\label{sec:qmc-observables}

A field acting on the Ising spins induces both Ising and source polarization. The ratio of these two responses measures the spin transferred to the source. To turn it into a contribution to the source's own susceptibility, we project source polarization onto the Ising mode using static thermal response as an inner product. Positivity makes the remaining source contribution nonnegative.

For $L^2$ cells, use the normalized modes
\begin{equation}
 \Phi_{\bm q}=\frac1L\sum_x e^{-i\bm q\cdot x}\tau_x^z,
 \qquad M_{a,\bm q}=\frac1L\sum_x e^{-i\bm q\cdot x}
 (n_{ax\uparrow}-n_{ax\downarrow}).
\end{equation}
Their connected response is
\begin{equation}
 \chi_{AB}(\bm q,i\Omega_n)=\int_0^\beta d\tau\,
 e^{i\Omega_n\tau}\langle A_{\bm q}(\tau)B_{-\bm q}(0)\rangle_c,
 \qquad \Omega_n=2\pi n/\beta.
 \label{eq:qmc-chi}
\end{equation}
The numerical estimator replaces the integral by $\Delta\tau\sum_\ell$, with both operators inserted at the left endpoint of the time step. The static data average the two directions $(2\pi/L,0)$ and $(0,2\pi/L)$ with the same normalization in every response. We write this average as $q_{\min}$. The cosine and sine spatial channels are Hermitian; their averaged response matrix is positive semidefinite. For the source and Ising polarizations this gives
\begin{equation}
 \chi_{aa}\chi_{\Phi\Phi}\ge |\chi_{a\Phi}|^2,
 \qquad R_a=\frac{\chi_{a\Phi}}{\chi_{\Phi\Phi}},
 \qquad \chi_{aa}\ge |R_a|^2\chi_{\Phi\Phi}.
 \label{eq:qmc-schwarz}
\end{equation}
The inequality already gives a lower bound on the source susceptibility. The projection identifies which source contribution the bound measures. For centered observables, static thermal response defines the positive Hermitian form
\begin{equation}
 (A,B)_\chi=\int_0^\beta d\tau\,\langle A(\tau)B^\dagger\rangle_c.
\end{equation}
The source observable $M_a^\perp=M_a-R_a\Phi$ is orthogonal to $\Phi$ in this form. Expanding its norm gives the exact decomposition
\begin{equation}
 \chi_{aa}=\chi_{\parallel}+\chi_{\perp},\qquad
 \chi_{\parallel}=|R_a|^2\chi_{\Phi\Phi},\qquad
 \chi_{\perp}=(M_a^\perp,M_a^\perp)_\chi\ge0.
 \label{eq:qmc-projection}
\end{equation}
Thus the measured mixed response determines a definite part of the source response itself: the part correlated with the collective Ising polarization. The residual contains every response orthogonal to that polarization. For the $q_{\min}$ average, this construction uses the direct sum of the two spatial directions and their Hermitian sine and cosine channels, with the averaging weights in the inner product. All terms in Eq.~\eqref{eq:qmc-projection} use the same ensemble, momentum, and static limit.

To place this source response in the spin system's ordering region, we measure the Binder cumulant and the spatial correlation length:
\begin{align}
 U_4&=1-\frac{\langle m^4\rangle}{3\langle m^2\rangle^2},
 &m&=\frac1{\beta L^2}\int_0^\beta d\tau\sum_x\tau_x^z(\tau),\\
 \frac{\xi_s}{L}&=\frac{1}{2L\sin(\pi/L)}
 \sqrt{\frac{\chi_{\Phi\Phi}(0,0)}{\chi_{\Phi\Phi}(q_{\min},0)}-1}.
 \label{eq:qmc-length}
\end{align}
Figure~3(a) of the Letter compares the measured correlation-length ratios across five fields for $L=4,6,8,10$. These curves identify the finite-size ordering region; the fixed-field responses in panels (b) and (c) follow the collective component over the same sizes.

\subsection{Hamiltonian and thermal ensemble}

We simulate the five-orbital Hamiltonian of Eqs.~\eqref{eq:collective-h}--\eqref{eq:collective-lattice}, with both fermion species and one quantum Ising spin in each cell. The contacts are all $J$, $t_s=\epsilon_s=0.05J$, $J_\tau=0.02J$, and $\lambda=0.1J$. These parameters are fixed throughout the size sequence. Electrons are grand canonical at the common side-orbital resonance. The spatial boundary conditions are periodic, and the principal sequence has $\beta J=50L$. The largest lattice shown in Fig.~3 has $10\times10$ cells, containing 500 spinful fermion orbitals and 100 quantum Ising spins.

The source, connector, and side orbitals remain quadratic, so the underlying Hamiltonian satisfies the support conditions of the particle-spectrum theorem in Sec.~\ref{sec:proof}. The Monte Carlo calculation below determines the collective response on this same source orbital. A separate imaginary-frequency comparison of the two responses is given in the companion.

For a real Ising path $\phi_x(\tau)=\pm1$, the two fermion species see opposite target potentials. The field scan shown here covers $L=4,6,8,10$ and $h_\tau/J=0.0605,0.06075,0.061,0.06125,0.0615$, with the common time step $\Delta\tau J=0.0625$. Figure~3 of the Letter uses this uniform grid. The conditional fermion step is $e^{-\Delta\tau h_0}e^{-\sigma\Delta\tau\lambda P_r^\dagger\phi P_r}$, where $\sigma=\pm1$ and $P_r$ selects target orbitals. The transverse-field spin transfer matrix is retained exactly at each step. The resulting split measure converges to the Hamiltonian ensemble as $\Delta\tau\to0$.

\subsection{Sampling the full fermion weight}

The sampled Ising configurations must include the feedback of the mobile electrons. We therefore accept cluster proposals for the imaginary-time Ising configuration~\cite{wolff1989} with the ratio of the two-species fermion determinants. Determinant sampling of quantum Ising fields is also used in studies of metallic criticality~\cite{schattner2016}. Here the contact structure makes the fermion weight positive. For continuous imaginary time, the free target covariance satisfies $\|P_r(\partial_\tau+h_0)^{-1}P_r^\dagger\|\le(2J)^{-1}$. Consequently $|\lambda|<2J$ excludes an antiperiodic zero mode for either species. Each real determinant stays positive as the coupling is increased continuously from zero.

The split measure has a separate finite-step criterion. Its free space-time Green matrix is $G_0=(I-Q)^{-1}$, where $Q$ propagates one time slice by $e^{-\Delta\tau h_0}$ and carries a minus sign across the time boundary. This sign imposes fermionic antiperiodicity. With $\Phi=\operatorname{diag}\phi$, define the dimensionless target kernel
\begin{equation}
 B=\tanh(\Delta\tau\lambda/2)\,[2G_0-I]_{rr}.
 \label{eq:qmc-kernel}
\end{equation}
Combining the two species gives the path-dependent weight $\mathcal W_f(\Phi)=C\det[I-(B\Phi)^2]$, with a positive path-independent constant $C$. The sampled parameters satisfy $\|B\|<1$, so the determinant is positive for every spin path. The series for $\operatorname{Tr}\log[I-(B\Phi)^2]$ gives bounded decision intervals. Ambiguous accept/reject decisions use the full stabilized determinant, preserving the original Metropolis rule. Source densities are evaluated deterministically with two terms of the convergent inverse series. The omitted terms and the finite convolution window have explicit bounds; the total per-density error budget is $10^{-8}$, including a numerical allowance of $10^{-10}$.

At every size and field, each coupling $\lambda=0,0.1J$ has independent ordered and disordered initial configurations. Each chain uses $10^5$ pure-Ising preparation clusters and 4000 full determinant-weighted warmup proposals, followed by 16000 production proposals for $L=4,6,8$ and 65536 for $L=10$. The $L=10$ data comprise twenty independent chains spanning the five fields, two couplings, and two initial conditions. Their disordered starts draw an independent periodic one-dimensional Ising path in imaginary time at each spatial site, incorporating the transverse-field temporal correlations before equilibration. Every proposal contains a fixed sequence of 16 Wolff clusters. The complete production interval enters the estimates. A spin measurement is saved after every production proposal, including rejected proposals. Source readouts, spaced by twenty proposals, evaluate the density on all sites at 16 uniformly selected imaginary-time slices. Uniform sampling without replacement gives an unbiased mixed-response estimator; its additional variance remains in the stored observations and error bars. The source ratio and projection use the Ising response on these same paired measurements; the denser Ising estimate is reported separately.

The thermal-response form used in Eq.~\eqref{eq:qmc-schwarz} also stays positive at a finite time step. The many-body transfer step $T=YD$ is similar to the positive Hermitian matrix $S=D^{1/2}YD^{1/2}$, where $Y$ contains kinetic hopping and the transverse spin field and $D$ contains the diagonal spin interactions and target potentials. Both measured densities commute with $D$. Their left-endpoint insertions therefore remain Hermitian under this similarity transformation, and the discrete thermal spectral representation has nonnegative coefficients. This supplies Eq.~\eqref{eq:qmc-schwarz} also for the sampled split ensemble.

\subsection{Predicting the transferred response from the free bands}
\label{sec:qmc-free-transfer}

Virtual polarization predicts the source--Ising overlap from the free
electronic bands. To compare this prediction with the measured mixed
response, we evaluate it at the same momentum, temperature, and time
step as the simulation. Let $M=\beta/\Delta\tau$ be the number of time
slices and define the free space--time covariance
\begin{equation}
 D_0(\bm k,\theta_m)=
 [I-e^{-i\theta_m}e^{-\Delta\tau h_0(\bm k)}]^{-1},
 \qquad \theta_m=(2m+1)\pi/M.
 \label{eq:split-free-covariance}
\end{equation}
This is the left-endpoint convention of the sampled transfer step.
The coefficient for a mode $Q=(\bm q,i\Omega_n)$ is
\begin{align}
 b_{a,\Delta\tau}(Q)&=2\lambda\kappa_{L,\beta,\Delta\tau}(Q),\nonumber\\
 \kappa_{L,\beta,\Delta\tau}(Q)&=
 \frac{\Delta\tau}{ML^2}\sum_{\bm k,m}
 [D_0(\bm k+\bm q,\theta_m+\Omega_n\Delta\tau)]_{ar}
 [D_0(\bm k,\theta_m)]_{ra}.
 \label{eq:split-polarization}
\end{align}
Here $\Omega_n=2\pi n/\beta$. Differentiating the conditional source
density with respect to the target potential gives this expression.
The two fermion species then give
$\chi_{a\Phi}(Q)/\chi_{\Phi\Phi}(Q)=b_{a,\Delta\tau}(Q)
+O(\lambda^3)$ at fixed $L,\beta,\Delta\tau$.
The coefficient $b_{a,\Delta\tau}$ depends on the stated lattice size and temperature. The reference contains no fitted parameter. Figure 3(c) of the Letter compares $\chi_{\parallel}$ with $b_{a,\Delta\tau}^2\chi_{\Phi\Phi}$, using the paired Ising response for both quantities. In the limit
$\Delta\tau\to0$ it approaches the finite-lattice free Kubo response;
the long-wavelength, zero-temperature, infinite-lattice limit gives
$2\lambda\kappa_a$ used in the analytic discussion.

\begin{center}
\begin{tabular}{cc}
\toprule
$L$ & $10^3 b_{a,\Delta\tau}$\\
\midrule
4 & $-1.915$\\
6 & $-1.910$\\
8 & $-1.908$\\
10 & $-1.908$\\
\bottomrule
\end{tabular}
\end{center}
These values use $\lambda=0.1J$, $\beta J=50L$, $\Delta\tau J=0.0625$,
and $\bm q=(2\pi/L,0)$; lattice symmetry gives the same result in the
other minimum-momentum direction.

\subsection{The uncoupled Ising reference}
\label{sec:qmc-baseline}

The uncoupled reference separates the collective fluctuations already present in the Ising system from the response transferred to the source by $\lambda$. We sample $\lambda=0$ on the same size, field and time-step grid as the coupled scan, at $\beta J=50L$. Each point uses the same preparation, 4000 warmup proposals, and size-dependent production length as its coupled counterpart. The two fermion species then have identical conditional densities, so the mixed source--Ising response vanishes exactly. The source retains its free spin susceptibility.

At $h_\tau=0.061J$ and $\Delta\tau J=0.0625$, the uncoupled correlation-length ratios for $L=4,6,8,10$ are $0.767(3)$, $0.772(3)$, $0.770(4)$, and $0.7567(18)$. Table~\ref{tab:qmc-baseline} compares the collective response with the coupled model at this same step. Both systems show collective growth. Switching on $\lambda$ transfers part of that motion to the source, producing the nonzero response in Fig.~3(c).

\begin{table}[htbp]
\centering
\caption{Matched collective responses $J\chi_{\Phi\Phi}(q_{\min},0)$ at $h_\tau=0.061J$, $\beta J=50L$, and $\Delta\tau J=0.0625$. Errors are one blocked-bootstrap standard error; coupled entries reproduce Fig.~3(b) of the Letter. At $\lambda=0$ the source--Ising mixed response is exactly zero; its nonzero value at $\lambda=0.1J$ gives the source contribution shown in Fig.~3(c).}
\label{tab:qmc-baseline}
\begin{tabular}{ccc}
\toprule
$L$ & $\lambda=0$ & $\lambda=0.1J$\\
\midrule
4 & $19.62(8)$ & $19.53(7)$\\
6 & $39.36(17)$ & $39.28(17)$\\
8 & $66.95(28)$ & $66.96(29)$\\
10 & $103.11(23)$ & $102.26(23)$\\
\bottomrule
\end{tabular}
\end{table}

\subsection{Predicting the response on a larger lattice}
\label{sec:qmc-size-prediction}

The $L=10$ lattice also tests whether the smaller lattices predict the evolution of the spin correlations. We determine the field dependence from $L=4,6,8$ and fix the resulting predictions before comparing them with the independent $L=10$ data. The prediction concerns all five fields and three observables, $\xi_s/L$, $U_4$, and $J\chi_{\Phi\Phi}$, together.

We use the short-range Ising scaling form with its leading size correction~\cite{wegner1972,kos2016}. For each dimensionless observable $Q=\xi_s/L,U_4$,
\begin{equation}
 Q(L,h_\tau)=A_Q+B_Qx+C_Qx^2+L^{-\omega}(D_Q+E_Qx),
 \qquad x=\frac{h_\tau-h_c}{J_\tau}L^{1/\nu}.
 \label{eq:qmc-size-prediction}
\end{equation}
The susceptibility has the same polynomial form multiplied by $L^{2-\eta}$, together with a size-independent regular background. The exponents are fixed to $\nu=0.62997$, $\eta=0.03630$, and $\omega=0.83$. Each coupling has a common $h_c$ and observable-dependent amplitudes, giving seventeen parameters. The critical field is profiled over the fixed domain $0.058\le h_c/J\le0.064$.

Within this specified scaling family and search domain, the independent $L=10$ responses are compatible with the predictions. The joint fifteen-component residual includes the covariance of all predicted and measured quantities. Its bootstrap $p$ values are $0.934$ for $\lambda=0$ and $0.273$ for $\lambda=0.1J$; Holm adjustment for the two comparisons gives $0.934$ and $0.546$. Thus the larger lattice supports the size evolution underlying Fig.~3, while the source response is predicted separately from the free-band polarization in Eq.~\eqref{eq:split-polarization}.

\subsection{Statistical precision}
\label{sec:qmc-statistics}

The source projection combines ratios and products of correlated ensemble moments. Its uncertainty therefore depends on both autocorrelation within a chain and covariance between the measured quantities. Joint resampling retains these correlations in the uncertainties of the response ratio and projection.

All ratios are formed after ensemble averaging. Independent chains are resampled independently, with their numerator, denominator, and other primitive moments kept together. For the uniform $\Delta\tau J=0.0625$ grid, we use 4000 joint circular moving-block bootstrap draws and retain every production observation. The primary block length is the next power of two above eight times the largest estimated correlation time, including the influence sequences of ratios and the source projection. The displayed grid uncertainties use this automatic block choice. Comparisons with block lengths 64, 128, 256, and 512 proposals accompany the data; the longer $L=10$ chains also use blocks of 1024 and 2048 proposals. Their spin-observable correlation times range from 0.53 to 1.15 proposals, and their error estimates reach a plateau over blocks of 256--2048 proposals. Ordered and disordered starts, fixed halves, and fixed quarters are compared using the complete production records.


\section{How the contact changes electronic feedback on the transition}
\label{sec:collective-feedback}

Transferring a spin signal to a metal can also change the spin dynamics being measured. In an ordinary metal, a spin fluctuation decays into low-energy particle--hole pairs. We first show how the target node suppresses this decay. We then control the longer-range interactions generated at every order and state the renormalization-group input that connects this microscopic result to a surviving Ising transition.

The result is a convergent, short-range induced spin action whose strength is $O(\lambda^2)$. With the smooth Ising matching and stability input stated in Sec.~\ref{sec:collective-stability}, one transverse-field tuning then restores the critical point. The calculation proceeds from the damping kernel to the full induced action and finally to that tuning. A bypass and one-side contacts provide physical controls by changing the target zero.

\subsection{The leading damping kernel}
\label{sec:collective-damping}

For a transparent coefficient, first linearize a circular Fermi surface with velocity $v_F$ and density of states $\nu_F$ per spin species. At zero temperature, only energies $-\omega<\epsilon<0$ can absorb a positive frequency $\omega$. The two target weights give
\begin{align}
 \Pi_\phi''(q,\omega)
 &=\frac{2\lambda^2\nu_F C_r^2}
 {\sqrt{(v_Fq)^2-\omega^2}}
 \int_{-\omega}^{0}\epsilon^4(\epsilon+\omega)^4\,\dd\epsilon\nonumber\\
 &=\frac{2\lambda^2\nu_F C_r^2}{630}
 \frac{\omega^9}{\sqrt{(v_Fq)^2-\omega^2}},
 \qquad 0<\omega<v_Fq.
 \label{eq:collective-damping}
\end{align}
The factor two counts the two spin species. At $\omega\ll v_Fq$, the familiar metallic $\omega/q$ absorption becomes $\omega^9/q$. On a scaling ray with $\omega/q$ fixed, this kernel has degree eight. A smooth anisotropic Fermi surface changes the angular function but preserves these eight additional energy powers.

To use this result in the imaginary-time spin action, we need the corresponding Matsubara response, including its real analytic terms. Let $\chi_{rr}$ denote the target-density response of one free species; its absorptive part gives $\Pi_\phi^{\prime\prime}=2\lambda^2\chi_{rr}^{\prime\prime}$. Put $a=v_Fq$ and $\nu=|\Omega_n|$. The leading low-band contribution is
\begin{equation}
 \chi_{rr}(q,\ii\nu)=\frac{\nu_F C_r^2}{630}
 \left[\frac{35a^8}{128}-\frac{5\nu^2a^6}{16}
 +\frac{3\nu^4a^4}{8}-\frac{\nu^6a^2}{2}+\nu^8
 -\frac{\nu^9}{\sqrt{\nu^2+a^2}}\right].
 \label{eq:collective-matsubara}
\end{equation}
Angular averaging of $\delta^{10}/(\delta^2+\nu^2)$, with $\delta=a\cos\theta$, gives this expression. The polynomial terms cancel the apparent $\nu^8$ remainder at $q=0$. They therefore must be kept when comparing static and dynamic limits. Contributions involving the other bands are analytic at low frequency and momentum and shift local spin parameters.

This leading calculation identifies the suppression mechanism. The effect on the transition also depends on interactions generated when high-energy degrees of freedom are eliminated. In particular, a coupling between the Ising energy density and a fermion density can destabilize an Ising critical point even when the direct order-parameter coupling is absent~\cite{nandkishore2012}. We therefore follow the target-amplitude factors through these generated interactions.

\subsection{Generated vertices keep the target-amplitude factors}
\label{sec:collective-factorization}

To see why generated couplings retain the zeros, separate the target field into its low- and high-band parts, $r=F\psi+U_h\chi$. Every low-band line reaches the interaction through its target component $F$. For any time-dependent spin trajectory, integration over the high-band fermions is Gaussian, and the resulting Schur complement keeps these factors explicit:
\begin{align}
 \mathcal G_l^{-1}[\phi]&=G_l^{-1}+F^\dagger\mathcal B[\phi]F,
 \label{eq:collective-factorization}\\
 \mathcal B[\phi]&=\lambda\phi-\lambda^2\phi U_h
 (G_h^{-1}+\lambda U_h^\dagger\phi U_h)^{-1}
 U_h^\dagger\phi .
 \label{eq:collective-B}
\end{align}
These equations use a Euclidean plus-sign convention; the sign of $\lambda$ is opposite for the other fermion species. The determinant of the high block produces an additional pure-spin action.

For $|\lambda|<\Delta_h$, the high-band inverse has a convergent expansion in $\lambda/\Delta_h$. Its coefficients are short-range kernels, with the factor $F$ on each external low-band fermion leg. When products of Ising fields combine into the Ising energy operator under coarse graining, these factors remain on the fermion legs. Contracting internal low-band lines also preserves the factors on the remaining external lines. Thus the generated energy-density coupling carries the same target-amplitude zeros as the direct vertex.

The orbital zeros therefore survive band elimination: every generated coupling to a low-band fermion retains the same suppressed target overlap. The next step determines how far these interactions can carry a spin fluctuation.


\subsection{Closed fermion cycles are short ranged to all orders}
\label{sec:collective-closedcycles}
\label{sec:collective-locality}

The important question for criticality is how far the electrons couple spin fluctuations. A particle can travel far through a metal, but an induced spin interaction contains a closed electron path: it must travel out and return. Both long parts of that path are suppressed by the target node. Bounding them together gives the range estimate below and allows us to sum the interactions over all orders.

Let $G_r(X)$ be the free target propagator in space and imaginary time,
with $X=(\tau,\bm x)$. Time is measured in fixed inverse contact-energy
units. More generally, suppose the target amplitude on an isolated low
band vanishes as $e_{\bm k}^{\ell}$ along its smooth Fermi contour. Its
contribution to the momentum-space propagator is
\begin{equation}
 F_0(\ii\omega,\bm k)
 =\frac{e_{\bm k}^{2\ell}a(\bm k)}{\ii\omega-e_{\bm k}}
 +F_h(\ii\omega,\bm k),
 \label{eq:rg9-lowprop}
\end{equation}
where $a$ is smooth and $F_h$ contains the gapped bands. The numerator
softens the low-band singularity at the Fermi contour. We quantify this
smoothness in momentum and frequency because it determines the decay of
the real-space propagator. Near the Fermi contour, the two transverse variables are $(\omega,e_{\bm k})$. On a shell of radius $r$, the propagator scales as $r^{2\ell-1}$; taking $s$ derivatives costs $r^{-s}$. The square-integrated shell contribution therefore scales as $r^{2(2\ell-s)}$. Summing shells down to zero is finite for $s<2\ell$: this is the Sobolev regularity statement $F_0\in H^s$.

Fourier transformation turns these derivatives into distance weights. Plancherel then gives a finite squared-propagator moment for every weight exponent $\alpha=2s<4\ell$. In two space dimensions, choosing $s>3/2$ also makes the remaining inverse weight integrable over space and time, so weighted Cauchy--Schwarz bounds the absolute propagator integral. Thus, for $\ell\ge1$, we obtain
\begin{align}
 M_0&=\int\dd X\,|G_r(X)|<\infty,\nonumber\\
 \mathcal I_\alpha&=\int\dd X\,(1+|X|)^\alpha|G_r(X)|^2<\infty,
 \qquad 0\le\alpha<4\ell.
 \label{eq:rg9-propagatornorms}
\end{align}
The first norm controls the total absolute propagation through the target;
the weighted second norm controls its long-distance tail. Here $\ell$ denotes the target-amplitude order; the complete-spectrum
contact order $p$ in Sec.~\ref{sec:unified} is a separate condition.
For our quadratic target zero, $\ell=2$, we may choose $s=3$ and hence $\alpha=6$. This is why a sixth distance moment is available for the later criticality argument. Here and below the integral includes the spatial lattice sum. The smooth
Fermi contour needs a nonzero velocity, but need not have nonzero
curvature everywhere.

At a fixed spin trajectory, the electron Hamiltonian is quadratic. Integrating out the electrons gives a fermion determinant; its logarithm expands into closed paths joining the spin insertions. The term of order $n$ contains $n$ vertices. Bounding one general cycle therefore controls each induced interaction before we sum over orders. Consider an $n$-vertex cycle, $n\ge2$, with $X_1=0$, $X_{n+1}=X_1$, and diameter
$D=\max_{i,j}|X_i-X_j|$. Its anchored kernel obeys
\begin{equation}
 \int\prod_{j=2}^n\dd X_j\,(1+D)^\alpha
 \prod_{j=1}^n|G_r(X_j-X_{j+1})|
 \le C_\alpha n^{\alpha+2}\mathcal I_\alpha M_0^{n-2}.
 \label{eq:rg9-cycle}
\end{equation}
The two long edges allow the diameter weight to be shared. Write
$f(X)=|G_r(X)|$, $w(X)=1+|X|$, and $Y_j=X_j-X_{j+1}$.
Choose vertices realizing $D$. The two arcs joining them contain
$m$ and $n-m$ edges. By the triangle inequality, there are distinct
edges $i,j$ on these arcs with $|Y_i|\ge D/m$ and
$|Y_j|\ge D/(n-m)$. Hence
\begin{align}
 (1+D)^\alpha
 &\le [m(n-m)]^{\alpha/2}
       w(Y_i)^{\alpha/2}w(Y_j)^{\alpha/2}\nonumber\\
 &\le (n/2)^\alpha
       w(Y_i)^{\alpha/2}w(Y_j)^{\alpha/2}.
 \label{eq:cycle-weight-split}
\end{align}
We can now sum over all $\binom n2$ possible edge pairs and drop
which-pair restrictions from each positive integral. Define
$f_\alpha=w^{\alpha/2}f$, so that
$\|f_\alpha\|_2^2=\mathcal I_\alpha$. Closure imposes
$\sum_jY_j=0$; every anchored edge integral is therefore a convolution
at zero. Young's inequality and Cauchy--Schwarz give
\begin{align}
 (f_\alpha*f_\alpha*f^{*(n-2)})(0)
 &\le \|f_\alpha*f_\alpha\|_\infty M_0^{n-2}\nonumber\\
 &\le \mathcal I_\alpha M_0^{n-2}.
 \label{eq:cycle-young}
\end{align}
The same estimate applies if an edge orientation reflects a factor.
For $n=2$, the convolution contains just the two weighted factors.
Combining these steps gives the sharper prefactor
$(n/2)^\alpha\binom n2$, and thus Eq.~\eqref{eq:rg9-cycle}
with $C_\alpha=2^{-\alpha-1}$. A sixth diameter moment uses
$\|w^3G_r\|_2^2=\mathcal I_6$, which is finite for $\ell=2$.
The two long edges of the closed path share the distance weight,
giving this six-power range estimate.

To include every induced spin interaction, we sum these cycles through the
fermion determinant. For a spin trajectory $\phi_i(\tau)=\pm1$, the two
species produce the exact fermion action
\begin{equation}
 \Delta S_f[\phi]
 =-\sum_{s=\pm1}\Tr\log(1+s\lambda F_0\phi)
 =\sum_{\substack{n\ge2\\n\ \mathrm{even}}}
 \frac{2\lambda^n}{n}\Tr(F_0\phi)^n.
 \label{eq:rg9-det}
\end{equation}
The exact side-orbital equation gives
$\|F_0(\ii\omega)\|\le
\min\{|\omega|/s_2^2,1/|\omega|\}\le1/|s_2|$.
The expansion therefore has a volume-independent convergence radius
$|\lambda|<|s_2|$. Its total action is extensive; convergence is uniform
for the action density. In particular, at zero temperature
\begin{equation}
 \frac{\|F_0\|_{\rm HS}^2}{\beta N}
 \le\int\frac{\dd\omega}{2\pi}
 \min\{\omega^2/s_2^4,1/\omega^2\}
 =\frac{4}{3\pi|s_2|}
 \label{eq:rg9-density}
\end{equation}
per target orbital, with a uniformly bounded Matsubara counterpart.

We have now bounded both the coefficient and the spatial extent of every cycle. Define $\|\Delta S_f\|_\alpha$ by summing their absolute kernels, fixing one vertex at the origin and weighting the remaining integral by $(1+D)^\alpha$. Inserting the determinant coefficient into Eq.~\eqref{eq:rg9-cycle} gives
\begin{equation}
 \|\Delta S_f\|_\alpha
 \le2C_\alpha\lambda^2\mathcal I_\alpha
 \sum_{\substack{n\ge2\\n\ \mathrm{even}}}
 n^{\alpha+1}(|\lambda|M_0)^{n-2}.
 \label{eq:rg9-actionsum}
\end{equation}
It is finite and of order $\lambda^2$ when $|\lambda|M_0<1$.
The polynomial dependence on cycle length preserves this nonzero
convergence radius after summing all vertex orders. For the two-side
contact, $\ell=2$: every absolute diameter moment below eight is finite.
We use the sixth moment below.

The same sixth-moment estimate holds uniformly on the finite-temperature, finite-volume systems used to approach criticality. Section~\ref{sec:uniform-limits} proves this step. We first use the bound to identify the generated interactions and the required critical tuning.

\subsection{Local terms, composite operators, and critical stability}
\label{sec:collective-rg9}
\label{sec:collective-stability}

Short range alone does not identify the effect of an induced interaction. Its spin content matters too. For example, two nearby Ising order fields contain a contribution from the local energy density. We therefore combine nearby insertions first, separate the changes of local Ising parameters, and then assess the interactions left between distant regions. This order of steps includes the energy channel along with the direct order-field channel.

\paragraph{Combine nearby spin insertions.}
Choose a fixed smooth momentum cutoff below the separated-band scale. In one Wilson blocking step, nearby insertions belong to the same coarse spatial and temporal block. Expand their product in local Ising operators, following a fixed order of grouping; this is the fusion prescription used below. Products
of microscopic spins are bounded. Their projections onto any finite
set of local operators have finite coefficients at this cutoff.
Replacing each cluster by a representative changes the diameter by
at most the block size, so the weighted kernel estimate survives a
bounded finite blocking step. More explicitly, for blocks of radius $b$, the representative
diameter obeys $D'\le D+2b$, so
$(1+D')^\alpha\le(1+2b)^\alpha(1+D)^\alpha$.
Choose a fixed finite operator projection whose retained coefficient
sum plus complementary remainder norm is at most $L\ge1$ per cluster.
Products of microscopic spins have norm at most one; an $n$-vertex
term therefore acquires at most $L^n$.
The cluster assignments partition the integration domain, so their
restricted integrals sum without an extra set-partition factor.
Identity clusters can be integrated out in the same absolute norm.
One remaining nonidentity cluster gives a local term; a nonlocal
remainder contains at least two. These are finite-step statements.
Iterating the blocking transformation at criticality requires the
RG input below.

The energy channel makes the physical consequence of this clustering
explicit. At fourth order, two nearby order operators can fuse into the
Ising energy operator $\varepsilon_{\rm I}$ in each of two distant regions.
A contribution to the resulting energy--energy
kernel has the form
\begin{align}
 K_{\varepsilon\varepsilon}(R)\supset{}&\lambda^4
 \int\dd r\,\dd s\,C_\varepsilon(r)C_\varepsilon(s)
 G_r(r)G_r(R+s-r)\nonumber\\[-2pt]
 &\hspace{26mm}\times G_r(-s)G_r(-R),
 \label{eq:rg9-energykernel}
\end{align}
where $r,s$ lie within the blocking region and $C_\varepsilon$ is the
local fusion coefficient. There are two long lines between the clusters. With $|r|,|s|\le b$,
$\Delta=s-r$, and $f=|G_r|$, submultiplicativity of $w$ gives
\begin{align}
 \int\dd R\,w(R)^\alpha f(R+\Delta)f(-R)
 &\le w(\Delta)^{\alpha/2}\mathcal I_\alpha\nonumber\\
 &\le(1+2b)^{\alpha/2}\mathcal I_\alpha.
 \label{eq:cycle-fused-moment}
\end{align}
The first inequality puts half the weight on each long factor and
uses Cauchy--Schwarz. If $|C_\varepsilon|\le L_\varepsilon$ in the
block, this contribution is bounded by
$|\lambda|^4L_\varepsilon^2(1+2b)^{\alpha/2}
\mathcal I_\alpha M_0^2$.
The other cyclic arrangements are covered by the cycle bound and
the representative-diameter inequality. The generated energy coupling
therefore has the same short-range moment bound, which excludes an
unsuppressed metallic $|\Omega|/q$ term. The high-band Schur complement supplies a
complementary check: every generated low-band bilinear retains the
external factors $F^\dagger$ and $F$, even when its spin operators fuse
into $\varepsilon_{\rm I}$.

\paragraph{Separate local shifts from long-range feedback.}
The same kernel contains analytic terms that change local Ising couplings. Its constant Fourier component shifts the mass, and powers of momentum give local derivative terms. We extract these pieces before asking whether a long-range remainder can change the critical behavior. The local subtraction retains both the constant mass shift and
derivative terms. Let $K$ be a fused kernel with $k$ nonidentity
clusters, and let $Q=(q_2,\ldots,q_k)$ denote its independent external
space--time momenta. Choose a smooth cutoff $\chi_\Lambda(Q)$ equal to
one near zero. For $m=\lfloor\alpha\rfloor$, define
\begin{equation}
 (\mathcal P_\alpha\widehat K)(Q)
 =\chi_\Lambda(Q)
 \sum_{|I|\le m}\frac{Q^I}{I!}\partial_Q^I\widehat K(0),
 \qquad
 \widehat R_K=\widehat K-\mathcal P_\alpha\widehat K .
 \label{eq:rg9-localization}
\end{equation}
The multi-index $I$ labels derivatives with respect to all components of $Q$, with $|I|$ their total order. The Taylor polynomial collects the local terms; the smooth cutoff confines this expansion to the low-momentum region. The remainder $\widehat R_K$ contains the nonlocal dependence whose scaling we must assess.
The absolute $\alpha$-moment makes these derivatives finite. Taylor's
integral remainder gives $\widehat R_K(Q)=O(|Q|^\alpha)$ for
noninteger $\alpha$ and $o(|Q|^\alpha)$ for integer $\alpha$, within
the region where $\chi_\Lambda=1$. Thus our sixth-moment choice includes
all local derivative terms through degree six; the one-side
third-moment comparison includes them through degree three. Choose the cutoff as a product of fixed smooth cutoffs in the
independent momenta. Its inverse Fourier transform is rapidly
decaying. At fixed derivative order six, the number of Taylor
terms grows only polynomially with the number of clusters.
The weighted norms of the subtracted terms therefore cost at most
$C^n n^r$ at original vertex order $n$, with fixed $C,r$.
They preserve the sixth-moment range class.

Apply this localization after each partial fusion in the chosen
nested blocking prescription. Treating each fused set of external
momenta in this way also controls the energy-channel clustering in
Eq.~\eqref{eq:rg9-energykernel}.

At the short-range three-dimensional Ising fixed point, the local
terms have a familiar classification. The identity shifts the free
energy. The exact joint Ising/species-exchange symmetry forbids a
longitudinal field. The energy operator is the one allowed relevant
scalar and is removed by tuning $h_\tau$. A constant stress-tensor
coupling changes the velocity or metric and is fixed by a coordinate
rescaling. Total derivatives do not contribute to a translation-invariant
bulk action. The remaining symmetry-allowed local operators are
irrelevant within the Ising stability basin. In particular, a local
product denoted $\varepsilon_{\rm I}^2$ decomposes into Ising operators
with their own scaling dimensions. Its relevant scalar contribution
is the energy operator already tuned above; $2\Delta_\varepsilon$
is the formal sum of the two factor dimensions.

After these local contributions are separated, the absolute
$\alpha$-moment bounds the integrated kernel beyond distance $L$ by
$CL^{-\alpha}$, while Eq.~\eqref{eq:rg9-localization} removes its
lower-order analytic part. On increasing the observation scale by a factor $b$, the overall translation center supplies $b^3$. The relative coordinates have already been integrated into the anchored kernel bound. The fields supply $b^{-\sum_j\Delta_{O_j}}$, and the nonlocal tail supplies at most $b^{-\alpha}$. Thus a remainder joining nonidentity operators $O_j$ has the linearized power-counting bound
\begin{equation}
 y\le3-\sum_j\Delta_{O_j}-\alpha.
 \label{eq:rg9-multicluster}
\end{equation}
For two order operators this is the familiar short-range versus long-range Ising criterion~\cite{sak1973}. They give the smallest allowed dimension sum.
Using $\Delta_\sigma\simeq0.5181$ and
$\Delta_\varepsilon\simeq1.413$~\cite{kos2016}, the sixth moment gives
\begin{equation}
 y_{\sigma\sigma}\le-4-\eta_{\rm I},\qquad
 y_{\varepsilon\varepsilon}\le-3-2\Delta_\varepsilon<-5.82.
 \label{eq:rg9-safeexponents}
\end{equation}
More clusters or energy operators have larger dimension sums and hence
smaller upper bounds. For a general kernel, the estimate bounds its
nonlocal scaling components; a definite RG eigenvalue applies when
the kernel is a scaling eigenfunction. Ordinary local Ising corrections
to scaling can decay more slowly.

\paragraph{From the microscopic bound to a critical interval.}
The microscopic estimates give an $O(\lambda^2)$ bound on the combined
strength and range of the induced interactions. To connect this result to a critical line, we use
the following RG stability assumption for retarded multipoint
interactions. It requires smooth matching to the Ising action and a
single symmetry-even direction that must be tuned to reach criticality.
The local fixed-point
description and nonlinear scaling fields are standard Wilson RG
inputs~\cite{wilson1971,wegner1972}; below we state their required
extension to this interaction class. A bound at each fixed order is not yet a small perturbation of the full action: all orders must be small in the same space. We therefore use one norm that controls both interaction range and the number of spin insertions.
Fix the ultraviolet scale and coordinate units, and give an anchored
even-spin kernel $\Phi_n$ the norm
\begin{equation}
 \|\Phi\|_{6,a}
 =\sum_{\substack{n\ge2\\n\ {\rm even}}}e^{an}
 \int\dd X_2\cdots\dd X_n\,(1+D)^6
 |\Phi_n(0,X_2,\ldots,X_n)|,\qquad a>0.
 \label{eq:rg11-norm}
\end{equation}
The factor $e^{an}$ leaves room for the growth in coefficients caused by
finite blocking and projection onto local operators. The kernels are
represented in ordered vertices; permutation
symmetrization is an average and cannot increase this norm.
Their completion in this norm is a weighted direct sum of anchored
kernel spaces.
The finitely many retained local scaling fields carry a
finite-dimensional norm. Equation~\eqref{eq:rg9-cycle} implies
\begin{equation}
 \|\Phi_\lambda\|_{6,a}\le
 2C_6e^{2a}\mathcal I_6\lambda^2
 \sum_{\substack{n\ge2\\n\ {\rm even}}}
 n^7(e^a|\lambda|M_0)^{n-2}
 \le K_{6,a}\lambda^2
 \label{eq:rg11-small}
\end{equation}
whenever $e^a|\lambda|M_0\le1/2$ and the determinant and high-band
expansions converge. Here $K_{6,a}<\infty$ is fixed independently of
$\lambda$, volume, and infrared cutoff. Thus the full perturbation approaches zero in a fixed norm,
including its vertex-order sum. To include finite blocking and
localization, first choose a stronger input weight $a_{\rm in}>a$ with
$a_{\rm in}-a>\log C$, increasing $C$ to include the factor $L$
above. Then $C^n n^r e^{an}\le C' e^{a_{\rm in}n}$.
Applying Eq.~\eqref{eq:rg11-small} with $a_{\rm in}$ absorbs these
finite-step costs by a smaller fixed coupling interval. The sixth
distance moment controls every step of this matching.

\paragraph{The stability input and the single tuning.}
The remaining input concerns how the Ising critical point responds to this small action perturbation. Let $\mathcal Y$ be a normed neighborhood of the Ising action containing the retained local operators and the retarded multipoint kernels, with overall even Ising parity. Its complete normed-space structure supplies the Banach chart used for the RG maps below. A fused
term may have an odd number of clusters, as in $\sigma\sigma\varepsilon$.
The fixed finite connected matching map $\mathcal Q(h_\tau,\Phi)$ is
jointly continuously differentiable from a neighborhood in
$\mathbb R\times\{\|\cdot\|_{6,a_{\rm in}}<\infty\}$ into $\mathcal Y$.
This input includes the connected spin cumulants generated by coarse
graining; the determinant above has already controlled the fermion
cycles.

The Wilson map on $\mathcal Y$ is continuously differentiable.
After removing the identity and fixing the metric, its linearization
at the Ising fixed point
has bounded invariant relevant and stable projections, one symmetry-even
relevant direction with multiplier greater than one, and a stable part $D_s$
with $\|D_s^j\|\le C_s\vartheta^j$ for some $\vartheta<1$.
Both maps have uniform $C^1$ control on fixed neighborhoods as the
infrared regulators are removed, as do the spectral-splitting bounds.

Finally, changing $h_\tau$ must actually change the relevant Ising scaling field at the uncoupled transition. This is the transversality condition: the transverse-field family crosses, rather than runs tangent to, the local critical manifold. The spaces, activities,
and bounds in this input are fixed before taking $\lambda\to0$.
The microscopic estimates establish smallness in this domain.
Together with the assumed differentiability and hyperbolicity of
the nonlinear Wilson map, they yield the critical interval below.

Under this input there is $\lambda_*>0$ such that every fixed
$0<|\lambda|<\lambda_*$ has an Ising critical point after tuning
$h_\tau$, with
\begin{equation}
 h_{\tau,c}(\lambda)-h_{\tau,c}(0)=O(\lambda^2).
 \label{eq:rg11-critical}
\end{equation}
The reason one tuning suffices is that the assumed RG has a local
codimension-one stable manifold: one condition selects the actions that
flow to the Ising fixed point. Let $\mathfrak m$ be a smooth coordinate normal to it and
$\mathcal Q(h,\Phi)$ the fixed finite matching map.
At $h_0=h_{\tau,c}(0)$,
$\mathfrak m(\mathcal Q(h_0,0))=0$, while transversality gives
$\partial_h[\mathfrak m\circ\mathcal Q](h_0,0)\ne0$.
The implicit-function theorem supplies $h_c(\Phi)$ for
$\|\Phi\|_{6,a_{\rm in}}<\delta$, with $\delta>0$ fixed and
$|h_c(\Phi)-h_0|\le L\|\Phi\|_{6,a_{\rm in}}$.
Choosing $\lambda_*$ so that $K_{6,a_{\rm in}}\lambda_*^2<\delta$
proves Eq.~\eqref{eq:rg11-critical}. The interval is nonzero;
its endpoint depends on the unquantified stability-neighborhood size.


\subsection{Quantitative kernel control in the extended metal}
\label{sec:extended-kernel}

The preceding proof gives a nonzero interval of coupling in which the induced interactions are short ranged. We now put an explicit scale on their strength for the two-dimensional metal in Eq.~\eqref{eq:collective-lattice}. Two free-propagator integrals suffice: one bounds absolute propagation, and the other bounds a pair of propagators. Evaluating upper bounds on the entire lattice and time axis makes the resulting coupling interval independent of system size. For the free target propagator $G_r(X)$ of one species, write $\dd X=\dd\tau\sum_{\bm x\in\mathbb Z^2}$ and define
\begin{equation}
 M_0=\int\dd X\,|G_r(X)|,\qquad
 \mathcal I_0=\int\dd X\,|G_r(X)|^2.
 \label{eq:extended-norms}
\end{equation}
For the parameters already specified, our conservative bounds are
\begin{equation}
 M_0<2.746/J,\qquad \mathcal I_0<0.2419/J
 \quad (T=0).
 \label{eq:extended-numbers}
\end{equation}

The bounds cover the infinite lattice and the whole time axis. They follow by enclosing the free band energies and orbital weights, controlling spatial sums with momentum derivatives, and integrating a positive envelope for the low-band tail. The companion section \emph{Explicit infinite-lattice bounds for the induced interaction} gives these inequalities and the outward-rounded arithmetic in full.

\paragraph{All induced interactions and finite temperature.}
With the two propagator norms bounded, each additional vertex is
controlled by the same expansion parameter. For an anchored $n$-vertex cycle, two $L^2$ factors and $n-2$
$L^1$ factors give $\int|K_n|\le \mathcal I_0M_0^{n-2}$ by Young's inequality.
The two species cancel odd powers and contribute $2\lambda^n/n$
at even order. Hence the sum of absolute interaction strengths
per cell and unit imaginary time obeys
\begin{equation}
 \mathcal B_0(\lambda)\le
 \frac{\mathcal I_0}{M_0^2}
 \left[-\log\!\left(1-(\lambda M_0)^2\right)\right],
 \qquad |\lambda|M_0<1.
 \label{eq:extended-action}
\end{equation}
This sums every even-order induced interaction, including the polarization
bubble and higher-vertex terms.
The positive series on the right increases with each norm, so their
upper bounds may be inserted directly.

At finite inverse temperature $\beta$ the free covariance is the
antiperiodic image sum of its zero-temperature counterpart.
The triangle inequality therefore gives $M_0(\beta)\le M_0(\infty)$.
Eliminating the side orbital $d$ in the free target resolvent gives
$|F_0(\ii\omega,\bm k)|\le|\omega|/(\omega^2+s_2^2)$: the rest of the
free system has the same resolvent sign and can only increase the
imaginary part of the denominator. With $x=\beta|s_2|/2$,
\begin{align}
 \mathcal I_0(\beta)&\le\frac1\beta\sum_{\omega_n}
 \frac{\omega_n^2}{(\omega_n^2+s_2^2)^2}\nonumber\\
 &=\frac{\tanh x+x\,\operatorname{sech}^2x}{4|s_2|}
 \le\frac{3}{8|s_2|}.
 \label{eq:extended-finiteT}
\end{align}
The last step uses $\tanh x\le1$ and
$\cosh^2x\ge1+x^2\ge2x$.
Table~\ref{tab:extended-kernel} lists conservative action bounds.
At $\lambda=.1J$, the zero-temperature bound is $0.00252J$, compared with the bare Ising exchange $J_\tau=.02J$. The table quantifies the induced action over a range of couplings. Converting it into a critical-field displacement uses
the matching and stability input in Sec.~\ref{sec:collective-stability}.

\begin{table}[t]
\caption{Upper bounds for the extended two-dimensional metal.
The last column holds at every temperature. Displayed bounds are rounded upward.}
\label{tab:extended-kernel}
\centering
\begin{tabular}{cccc}
$\lambda/J$ & $|\lambda|M_0$ &
$10^3\mathcal B_0(0)/J$ & $10^3\sup_T\mathcal B_0(T)/J$\\
\hline
0.025 & 0.0687 & 0.152 & 0.235\\
0.05  & 0.138 & 0.611 & 0.947\\
0.1   & 0.275 & 2.52 & 3.90\\
0.2   & 0.550 & 11.6 & 17.9
\end{tabular}

\end{table}

The range estimate also allows an exponential vertex-order weight whenever $e^a|\lambda|M_0<1$. At $\lambda=.1J$, the choice $e^a=2$ gives a parameter below $.55$. The resulting microscopic bounds control the size and range of the input to the RG matching. The critical interval in Eq.~\eqref{eq:rg11-critical} additionally depends on the stability-neighborhood radius specified in Sec.~\ref{sec:collective-stability}; that radius is not determined by the propagator bounds.


\subsection{A bypass restores relevant metallic feedback}
\label{sec:collective-bypass}

A connectivity change provides a direct test of the mechanism. Add the hopping $w(a_i^\dagger d_i+\hc)$. At zero energy, the orbital equations give
\begin{equation}
 r=-\frac{w}{s_2}a,\qquad
 Z_w=\left[1+\frac{w^2}{s_2^2}
 +\frac{(g-vw/s_2)^2}{s_1^2}\right]^{-1}.
 \label{eq:collective-bypassweight}
\end{equation}
The target weight is now $Z_ww^2/s_2^2$ at the Fermi contour. The conventional damping coefficient is restored as
\begin{equation}
 \gamma_w\propto\lambda^2Z_w^2\frac{w^4}{s_2^4}.
 \label{eq:collective-bypassdamping}
\end{equation}
At the Ising fixed point this perturbation has eigenvalue $2-\eta_{\rm I}$. With fixed nonzero $\lambda$ and the mass retuned, its crossover scale therefore obeys
\begin{equation}
 \frac{\omega_*(w)}{\Lambda}
 \sim C_w\left|\frac{w}{J}\right|^{4/(2-\eta_{\rm I})},
 \qquad \frac{4}{2-\eta_{\rm I}}\simeq2.03697.
 \label{eq:collective-bypassscale}
\end{equation}
The constant $C_w$ depends on velocities, field normalization, and the fixed coupling. This scale marks the onset of relevant metallic feedback and departure from the protected fixed point.

A second comparison removes both side orbitals and keeps the $a$--$b$--$r$ chain. At unit contact couplings it has the same Fermi contour, source residue $1/2$, and Fermi velocity as Eq.~\eqref{eq:collective-h}, but target weight $1/2$ at the Fermi contour. The leading damping ratio is
\begin{equation}
 \frac{\Pi''_{\rm two\ sides}}{\Pi''_{\rm no\ sides}}
 \sim\frac{1}{630}\left(\frac{\omega}{J}\right)^8.
 \label{eq:collective-control}
\end{equation}
The comparison holds the low-energy metallic content fixed while changing its feedback on the spins. The two spin masses must be tuned separately because their analytic high-band corrections differ.

\subsection{One zero suppresses feedback; two zeros control the complete spectrum}
\label{sec:geometry-hierarchy-round9}

Suppressing metallic damping and protecting the complete particle spectrum are related design goals with different requirements. One target-amplitude zero already weakens decay into low-energy particle--hole pairs of the free metal. The complete spectral theorem also covers resonances created by an arbitrary interacting environment. Two zeros provide enough suppression for that stronger comparison while allowing increasingly fine relative resolution.

A useful comparison must keep a dispersive metallic band while removing one zero. Directly deleting $d$ from Eq.~\eqref{eq:collective-h} leaves a target-active zero-energy flat band; deleting $c$ gives a four-site chain with $\det h_0(0)=g^2s_2^2\ne0$. We therefore adjust the remaining contact parameters to retain a metal. With $c$ alone, giving $r$ a finite onsite energy $R\ne0$ produces an isolated low band with
\begin{equation}
 Z=\frac{s_1^2}{s_1^2+g^2},\qquad
 |u_r(e)|^2=\frac{Zg^2v^2}{s_1^4R^2}e^2+O(e^3).
 \label{eq:geometry-c-only}
\end{equation}
For a single transition-side leaf, the three-orbital contact
$h_d(\xi)=\left(\begin{smallmatrix}\xi&g&0\\g&0&s\\0&s&0\end{smallmatrix}\right)$
in the order $(a,r,d)$ has $Z=s^2/(s^2+g^2)$ and
$|u_r(e)|^2=Zg^2e^2/s^4+O(e^4)$. Setting $s=s_1$ matches the source residue, Fermi contour, and Fermi velocity of the two-side model. Retaining the original $a$--$b$--$r$ backbone is also possible: a connector onsite energy $\beta\ne0$ gives a crossing at $\xi=g^2/\beta$ and again a simple target zero.

The two responses require different information about the contact. Let $\ell$ be the clean-band amplitude order, $u_r(e)=O(e^\ell)$, and retain $p=m+n$ for the complete-spectrum criterion in Sec.~\ref{sec:unified}. The former describes the free band; the latter enters the bound on all many-body transitions. Their values coincide in these controls. In the damping calculation, each member of a particle--hole pair contributes its target weight. For a smooth circular Fermi surface with $|u_r(e)|^2=C e^{2\ell}$,
\begin{equation}
 \Pi''(q,\omega)\sim\frac{2\lambda^2\nu_F C^2}{v_Fq}
 \frac{[(2\ell)!]^2}{(4\ell+1)!}\,\omega^{4\ell+1}.
 \label{eq:geometry-damping-hierarchy}
\end{equation}
The leading damping and uniform spectral bounds thus compare as follows:
\begin{center}
\begin{tabular}{lcccc}
\toprule
Contact & $\ell$ & $(m,n)$ & Damping & Relative spectral parameter\\
\midrule
No zero & $0$ & $(0,0)$ & $\omega/q$ & $O(\eta^{-2})$\\
Propagation-side zero & $1$ & $(1,0)$ & $\omega^5/q$ & $O(E^2/\eta^2)$\\
Transition-side zero & $1$ & $(0,1)$ & $\omega^5/q$ & $O(E^2/\eta^2)$\\
Both & $2$ & $(1,1)$ & $\omega^9/q$ & $O(E^4/\eta^2)$\\
\bottomrule
\end{tabular}
\end{center}
The constants contain the fixed contact energies. A simple zero already suppresses Landau damping. Two independent zeros supply the additional environment-uniform resolving window $E^2/\varepsilon_g\ll\eta\ll E$.

\paragraph{Explicit limits of spectral protection with one zero.}
The following interacting examples produce an order-one change of the complete source spectrum with a single zero. In the first, a fixed interaction moves the target level into resonance. In fixed hopping units, use the $c$-only contact with $g=v=s_1=R=1$, source onsite $2\epsilon$, and an allowed density interaction $-n_rn_f$. In the stationary sector $n_f=1$, the interaction changes the target onsite from one to zero. The free source pole is $\epsilon+O(\epsilon^2)$ with weight $1/2+O(\epsilon)$; the interacting source-visible low pole is $4\epsilon/3+O(\epsilon^2)$. The remaining exact zero mode has zero source weight. Resolving the free pole at $\omega=\epsilon$, $\eta=\epsilon^{3/2}$ therefore gives
\begin{equation}
 A_0\sim\frac{1}{2\pi\eta},\qquad
 A_1=O(\eta/\epsilon^2),\qquad A_1/A_0\longrightarrow0.
 \label{eq:geometry-p1-counter}
\end{equation}
Here the interaction strength stays fixed. For the $d$-only three-orbital contact with $g=s=1$, source onsite $2\epsilon$, and interaction $\epsilon^{-1}n_rn_f$, the free low pole is $\epsilon+O(\epsilon^3)$ while the occupied-$f$ sector has poles $\pm\sqrt2\epsilon+O(\epsilon^3)$, giving the same spectral departure. This second sequence tests uniformity over interaction strength. Together the sequences rule out a complete-spectrum guarantee uniform over the allowed class with improving relative resolution from a simple zero alone. The response of an individual fixed environment remains model-dependent.


\subsection{Uniformity in temperature and volume}\label{sec:uniform-limits}
The range bound must remain finite as the system grows and the
temperature falls, since both limits are needed at a quantum critical
point. We establish this uniformity for the sixth moment. For the specified periodic contact, the low-band target
weight is $w_0(\bm k)=e_{\bm k}^4a(\bm k)$, with smooth band data through
third order. All other bands are separated from zero by a fixed gap.
At inverse temperature $\beta$, the target covariance is
\begin{equation}
 g_\beta(\tau,\bm k)
 =-\sum_b w_b(\bm k)
 \frac{e^{-\tau E_b(\bm k)}}{1+e^{-\beta E_b(\bm k)}},
 \qquad 0<\tau<\beta.
 \label{eq:rg13-thermal-covariance}
\end{equation}
Here $w_b$ is the target weight in band $b$. Degeneracies within a gapped
spectral subspace can be treated by matrix functional calculus.
Write $d_\beta(\tau)=\min(\tau,\beta-\tau)$ for the shortest distance around the imaginary-time circle. We need three momentum derivatives because their squared Fourier norm controls the sixth spatial moment. For each momentum multi-index $\nu$ with $j=|\nu|\le3$,
\begin{equation}
 \sup_{\bm k}|\partial_{\bm k}^{\nu}g_\beta(\tau,\bm k)|
 \le C_j(1+d_\beta(\tau))^{j-4},
 \label{eq:rg13-thermal-derivatives}
\end{equation}
with constants independent of $\beta$ and the spatial torus sizes.
To see this for the low band at $e\ge0$, differentiate
$e^4a e^{-\tau e}/(1+e^{-\beta e})$.
Each derivative reduces the fourth-order zero by at most one, or
produces a factor $\tau$ or $\beta$. Derivatives of the thermal factor
also supply $e^{-\beta e}$. Applying
$\sup_{e\ge0}e^m e^{-te}\le C_m t^{-m}$ to each term gives
$(1+\tau)^{j-4}$. At $e<0$, rewrite the thermal factor using
$\beta-\tau$ and repeat the same argument. Gapped contributions decay
exponentially in $d_\beta$ and obey the same bound.

Let $d_L(\bm x)$ be the length of the shortest coordinate representative
on a spatial torus. Parseval's identity gives
\begin{align}
 \sum_{\bm x}d_L(\bm x)^6|G_{\beta,L}(\tau,\bm x)|^2
 &\le C(1+d_\beta)^{-2},\nonumber\\
 \sum_{\bm x}|G_{\beta,L}(\tau,\bm x)|^2
 &\le C(1+d_\beta)^{-8}.
 \label{eq:rg13-spatial-moments}
\end{align}
In infinite space, the first bound uses the third momentum derivatives.
On a finite torus, use third forward differences with step
$h_i=2\pi/L_i$ instead. Their Fourier multipliers are
$[(e^{\ii h_ix_i}-1)/h_i]^3$, whose magnitudes are uniformly comparable
to $|x_i|^3$ for the shortest representatives. A finite difference is an
average of the corresponding derivative, so
Eq.~\eqref{eq:rg13-thermal-derivatives} applies with the same constants.

The spatial sixth moment therefore decays as $(1+d_\beta)^{-2}$. Multiplying the unweighted estimate by the temporal sixth power gives the same decay: $(1+d_\beta)^6(1+d_\beta)^{-8}=(1+d_\beta)^{-2}$. Both parts are integrable with a bound independent of the circle length. Using $|X|=d_\beta(\tau)+d_L(\bm x)$, we obtain
\begin{align}
 \mathcal I_6(\beta,L)
 &=\int_0^\beta\dd\tau\sum_{\bm x}
 (1+d_\beta+d_L)^6|G_{\beta,L}(\tau,\bm x)|^2\nonumber\\
 &\le C\int_0^\beta(1+d_\beta(\tau))^{-2}\dd\tau
 \le 2C.
 \label{eq:rg13-uniform-sixth}
\end{align}
Weighted Cauchy--Schwarz then bounds $M_0$ uniformly: the remaining
integral of $(1+d_\beta+d_L)^{-6}$ is bounded in three-dimensional
space--time. The absolute covariance is periodic even though the fermion
covariance is antiperiodic. Thus the two-edge convolution proof applies
on these tori, with their shortest-distance metric. The sixth-moment
action bound and its nonzero coupling interval are uniform as both
infrared regulators are removed.


\subsection{Relation to interference and impurity criticality}
\label{sec:collective-context}

Destructive interference can create zeros in the orbital weight seen by an impurity and can produce quantum-critical behavior in an interacting impurity problem~\cite{diasdasilva2006,diasdasilva2008,fritz2004}. Nodal interactions also support long-lived fermions and anomalous transport, including extended nodal surfaces~\cite{wang2025}. A critical metal with an entirely cold Fermi surface is possible through other mechanisms, such as coupling to lattice strain~\cite{paul2017}.

The collective application here asks how the metal feeds back on a spatially extended quantum order parameter. This feedback includes both the direct order-parameter vertex and generated couplings to the Ising energy density~\cite{nandkishore2012}. The exact orbital factors and the all-order range estimate control it within a fixed weak-coupling interval. Under the stated Ising stability input, virtual polarization gives the same source orbital a critical spin response. The general spectral theorem simultaneously controls its complete one-particle addition--removal spectrum. The electronic spectrum retains its free low-energy form while a collective response of those orbitals becomes singular.


\section{Measuring the collective response on physical orbitals}
\label{sec:collective-measurement}

Two measurements reveal the separation in the Letter: particle addition and removal determine the protected spectrum, while a spin-dependent field detects collective motion on the source. The spin measurement selects frequency and momentum to isolate the collective signal. An empty and an occupied auxiliary level supply the two channels of the particle measurement. Both protocols address the same physical source orbital.

\subsection{Source-resolved collective absorption}
\label{sec:source-dynamic-readout}

Drive the source spin with a weak, nearly uniform oscillating field and measure the absorbed power. This addresses the same physical orbitals as the particle probe. The relevant distinction is between spin on one orbital and total electron spin: virtual polarization can move spin among orbitals while conserving the total. A source-resolved drive can therefore excite the transferred collective motion.

Finite-frequency absorption isolates a small collective signal by selecting a window below the interband gap where the free source has no dissipative response. At zero coupling, the uniform source spin projected onto the low band is
\begin{equation}
 M_a(\bm q=0)=\sum_{\bm k}|u_a(\bm k)|^2
 (n_{\bm k\uparrow}-n_{\bm k\downarrow})+\text{interband terms}.
 \label{eq:source-uniform-projection}
\end{equation}
Every displayed occupation commutes with the free Hamiltonian. At zero temperature the free uniform source therefore has no absorption for $0<\omega<\Delta_h$. Its static Pauli susceptibility is a different limit. For small nonzero $q=|\bm q|$, the intraband continuum is confined to $\omega\le v_{\max}q+O(q^2)$, where $v_{\max}$ is the maximum low-band group velocity. A measurement in the window
\begin{equation}
 \max(v_{\max},c_{\rm I})q\ll\omega\ll\Delta_h,
 \qquad T\ll\omega,
 \label{eq:source-absorption-window}
\end{equation}
retains the critical Ising absorption while removing the free particle--hole background. A field approximately uniform over a region of size $\ell$ requires $\ell\gg\max(v_{\max},c_{\rm I})/\omega$ to resolve this window. At nonzero coupling, the leading source absorption is $b_a^2\chi_{\phi\phi}''$ together with higher-order regular corrections. Frequency and momentum resolution must be included when comparing these terms. The window removes the order-$\lambda^0$ dissipative background, while the absolute collective signal retains its coefficient $b_a^2$.

For a critical Ising sector, the positive-frequency continuum is
\begin{equation}
 \chi_{\phi\phi}''(q,\omega)
 =\mathcal C_\phi\sin(\pi\eta_{\rm I}/2)
 \Theta(\omega-c_{\rm I}|q|)
 (\omega^2-c_{\rm I}^2q^2)^{-1+\eta_{\rm I}/2},
 \label{eq:collective-criticalabsorption}
\end{equation}

Here $\mathcal C_\phi>0$ fixes the normalization of the order field and $c_{\rm I}$ is its velocity. The transferred source absorption has the same continuum multiplied by $b_a^2$.

A field uniform across cells measures the $q=0$ source response, which scales as $\chi_{aa}''(0,\omega)\propto b_a^2\omega^{-2+\eta_{\rm I}}$. A single-site probe sums over momenta. Integrating the critical continuum of Eq.~\eqref{eq:collective-criticalabsorption} over both momentum components gives
\begin{equation}
 \chi_{aa,\mathrm{loc}}''(\omega)
 =\frac{b_a^2\mathcal C_\phi\sin(\pi\eta_{\rm I}/2)}
 {2\pi\eta_{\rm I}c_{\rm I}^2}\,\omega^{\eta_{\rm I}}.
 \label{eq:source-local-absorption}
\end{equation}
The single-site local spin response is nonanalytic, but its static infrared contribution is finite for $\eta_{\rm I}>0$. The divergent collective response refers to orbital-resolved correlations across cells.

\subsection{Particle addition and removal by weak tunneling}
\label{sec:readout}

Weak tunneling measures the addition and removal spectra by exchanging a particle with an auxiliary level. It applies to any stationary host state and directly recovers the full spectral observable of the theorem. We derive the probe kernel and then bound the effects of finite pulse duration and coupling. The companion develops a finite-angle acquisition protocol for specified mixed-spin preparations.

A weak particle probe measures the spectrum through a physical source orbital. Transport through interacting regions generally also depends on occupation information~\cite{meir1992}. Here two probe preparations supply the addition and removal contributions explicitly. Local energy-resolved atomic probes provide an established spectroscopy setting~\cite{gruss2018}. For an explicit resolution kernel, take an auxiliary fermion $p$ with energy $\Omega_p$ and pulse
\begin{equation}
 H_p(t)=\lambda_p e^{-\eta t}(p^\dagger a_x+a_x^\dagger p),\qquad t\ge0.
 \label{eq:probe}
\end{equation}
Prepare the same stationary host state twice, with the probe initially empty in one experiment and occupied in the other. The transfer amplitude for a host transition of frequency $\epsilon$ contains
\begin{equation}
 \int_0^\infty\dd t\,e^{-\eta t}e^{\ii(\epsilon-\Omega_p)t}
 =\frac1{\eta-\ii(\epsilon-\Omega_p)}.
 \label{eq:pulse}
\end{equation}
The empty probe measures removal weight and the occupied probe measures addition weight. Summing their transfer probabilities gives
\begin{equation}
 P_{0\to1}+P_{1\to0}
 =\lambda_p^2\int\frac{\dd\mu_x^\rho(\epsilon)}{(\epsilon-\Omega_p)^2+\eta^2}
 +O\bigl((\lambda_p/\eta)^4\bigr)
 =\frac{\pi\lambda_p^2}{\eta}A_x^\rho(\Omega_p;\eta)
 +O\bigl((\lambda_p/\eta)^4\bigr).
 \label{eq:probability}
\end{equation}
The two preparations directly supply the canonical weights $p_m$ and $p_n$ in Eq.~\eqref{eq:measure} for any stationary host state. The time-integrated interaction-picture coupling has norm at most $\lambda_p/\eta$. Transfer amplitudes contain odd numbers of tunnel events, giving a quadratic leading probability and a fourth-order correction. The Dyson-series estimate is uniform in the host volume.

The inferred linear-response spectrum is therefore
\begin{equation}
 \lim_{\lambda_p/\eta\to0}\frac{\eta}{\pi\lambda_p^2}
 (P_{0\to1}+P_{1\to0})=A_x^\rho(\Omega_p;\eta),
 \label{eq:readoutlimit}
\end{equation}
with the weak-probe limit taken at fixed $\eta>0$. The relative theorem applies to this inferred quantity.

\paragraph{Finite pulse duration.}

The exponential pulse can be stopped at a finite time $T$. At leading order in the probe coupling, its spectral kernel is the infinite-time Lorentzian multiplied by
\begin{equation}
 \left|1-e^{-[\eta-\ii(\epsilon-\Omega_p)]T}\right|^2.
 \label{eq:finitepulse}
\end{equation}
The fractional change of this factor from one is at most
\begin{equation}
 \epsilon_T=2e^{-\eta T}+e^{-2\eta T}.
 \label{eq:pulseerror}
\end{equation}
Positivity of the host spectrum makes this a relative bound on the inferred intensity at every probe frequency, including weak signals. Let $\R_\Delta$ be the relative spectral bound for the actual contact parameters, derived in Sec.~\ref{sec:detuning}; it equals $\R$ for ideal resonant contacts. If $A_{x,T}$ denotes the spectrum inferred from the truncated pulse in the weak-probe limit, then
\begin{equation}
 \left|\frac{A_{x,T}}{A_{0,x}}-1\right|
 \le \R_\Delta+(1+\R_\Delta)\epsilon_T.
 \label{eq:combinedbudget}
\end{equation}
A sufficient duration for a $1\%$ pulse-truncation budget is
$T\ge5.31/\eta> -\log(\sqrt{1.01}-1)/\eta$.
Together with the $3.4\%$ interaction bound at the stated calibration tolerances in Sec.~\ref{sec:detuning}, this gives less than $4.5\%$ distortion. The finite-coupling estimate below completes this static readout bound.

\paragraph{Finite probe coupling.}

A finite-probe budget connects spectral protection to the measured intensity. Let $P=P_{0\to1}+P_{1\to0}$ be the exact transfer sum for a pulse of duration $T$, and let $Q$ be its second-order value. Define
\begin{equation}
 \kappa_T=\frac{|\lambda_p|}{\eta}(1-e^{-\eta T}),\qquad
 r_T=\sinh\kappa_T-\kappa_T.
\end{equation}
Each tunnel event reverses the probe occupation. The transfer block of the interaction-picture propagator therefore contains only odd Dyson orders; its difference from the first-order amplitude has norm at most $r_T$. For a mixed host state, average the squared amplitudes with its probabilities; equivalently, use the Hilbert--Schmidt norm weighted by that state. The bound $r_T$ still holds for each probe preparation. Place the empty- and occupied-probe amplitudes in two components of one vector. Its norm is $\sqrt P$, the first-order vector has norm $\sqrt Q$, and their difference has norm at most $\sqrt2r_T$. The triangle inequality gives $|\sqrt P-\sqrt Q|\le\sqrt2r_T$. Squaring this estimate yields
\begin{equation}
 \left|\frac P Q-1\right|
 \le b_p(Q,\kappa_T):=\frac{2\sqrt2 r_T}{\sqrt Q}+\frac{2r_T^2}{Q}.
 \label{eq:probebackaction}
\end{equation}
This finite-coupling relative bound is uniform in host volume and relates the allowed probe strength to the local signal. The reference spectrum fixes a conservative coupling in advance: for $\R_\Delta,\epsilon_T<1$,
\begin{equation}
 Q\ge Q_{\min}:=\frac{\pi\lambda_p^2}{\eta}
 (1-\epsilon_T)(1-\R_\Delta)A_{0,x}.
 \label{eq:Qmin}
\end{equation}
Choose $\lambda_p$ so that $b_p(Q_{\min},\kappa_T)\le b_*$. The finite-coupling mean inferred signal $\widehat A=\eta P/(\pi\lambda_p^2)$ then satisfies
\begin{equation}
 \left|\frac{\widehat A}{A_{0,x}}-1\right|
 \le(1+\R_\Delta)(1+\epsilon_T)(1+b_*)-1.
 \label{eq:totalreadout}
\end{equation}
The three factors respectively control interaction distortion, pulse truncation, and probe backaction, all relative to the same actual calibrated free reference.


\section{Calibration and finite windows with imperfect contacts}
\label{sec:implementation}

The result becomes an operating principle once its energy scales are tied to controllable device parameters. Bond amplitudes set the protection scale; side-level detuning changes its low-energy window; calibration fixes which free spectrum is being compared. Weak interactions on the source introduce a further scale because they can feed its collective polarization back into low-energy particle motion. We derive these effects separately and then combine their windows.

\subsection{Independent side-energy and bond errors}
\label{sec:detuning}

Independent side-energy and bond errors set a quantitative calibration window for the protected spectrum. Add $\sum_j(\Delta_{cj}n_{c_j}+\Delta_{dj}n_{d_j})$ to the Hamiltonian and compare the interacting spectrum with the free reference having these same actual onsite energies and couplings. Detuning shifts the two side-orbital factors from $z$ to $z-\Delta_c$ and $z-\Delta_d$. Let $D_c,D_d,S_1,S_2,V$ be real diagonal contact matrices, and include $D_c$ in the first-stage Hamiltonian $h_C$. With $A_c=zI-D_c$ and $A_d=zI-D_d$, the combined factor and its relative error parameter are
\begin{equation}
 B_\Delta(z)=A_d S_2^{-1}V S_1^{-1}A_c,\qquad
 \delta_\Delta=\frac{\|B_\Delta(z)\|^2}{\eta^2}
 =\max_j\frac{|v_j|^2|z-\Delta_{cj}|^2|z-\Delta_{dj}|^2}
 {|s_{1j}s_{2j}|^2\eta^2}.
 \label{eq:detunederror}
\end{equation}
The source hopping and source-contact matrices remain arbitrary. For $F_\alpha=G_{rr}^{(\alpha)}$ and $Q_\alpha=G_{dd}^{(\alpha)}$, the exact side equations now give
\begin{equation}
 S_2 F_\alpha S_2=A_dQ_\alpha A_d-A_d
 =A_d[Q_\alpha-A_d^{-1}]A_d.
 \label{eq:detunedleaf}
\end{equation}
Thus the common contact term cancels between environments. With $\widetilde u=(u,0,0)^T$ and $x_z=P_cg_C(z)\widetilde u$, the first-side equation gives $P_bg_C\widetilde u=S_1^{-1}A_c x_z$, hence
\begin{equation}
 u^\dagger(G_{aa}^{(1)}-G_{aa}^{(0)})u
 =[B_\Delta(\bar z)x_{\bar z}]^\dagger(Q_1-Q_0)
 B_\Delta(z)x_z.
 \label{eq:detunedpropagation}
\end{equation}
For a probe $u$, write $A_{\alpha,u}=u^\dagger\A_\alpha u$ and $A_{C,u}=u^\dagger\A_Cu$. The canonical bound $\|Q_1-Q_0\|\le1/\eta$, together with the free resolvent identity, gives
$|A_{1,u}-A_{0,u}|\le\delta_\Delta A_{C,u}$.
We must again replace the disconnected intensity $A_{C,u}$ by the measured free reference. Since $A_d^{-1}$ is itself a normalized canonical Green matrix, $\|Q_\alpha-A_d^{-1}\|\le1/\eta$; Eq.~\eqref{eq:detunedleaf} then gives $A_{0,u}\ge(1-\delta_\Delta)A_{C,u}$. Dividing the two estimates produces $\delta_\Delta/(1-\delta_\Delta)$. Since this holds for every $u$, we obtain
\begin{equation}
 (1-\R_\Delta)\A_0\preceq\A_1\preceq(1+\R_\Delta)\A_0,
 \qquad\R_\Delta=\frac{\delta_\Delta}{1-\delta_\Delta},\quad\delta_\Delta<1.
 \label{eq:detunedbound}
\end{equation}
This remains uniform over the stationary state and the allowed environment. It controls interaction-induced changes relative to the actual calibrated free network. Changes of that free network relative to an ideal design are a separate one-particle calibration problem.

For a common calibration origin $\Omega$, write $E=|\omega-\Omega|$ and suppose $|\Delta_{cj}-\Omega|,|\Delta_{dj}-\Omega|\le D$. A directly usable envelope is
\begin{equation}
 \delta_\Delta\le\frac{[(E+D)^2+\eta^2]^2}{\varepsilon_g^2\eta^2}.
 \label{eq:calibrationenvelope}
\end{equation}
Relative to the ideal envelope, the factor is $\{[(E+D)^2+\eta^2]/[E^2+\eta^2]\}^2$. Thus physical side-energy offsets $D=O(\eta)$ change it by $1+O(\eta/E)$ when $\eta/E\to0$. This differs from uncertainty in a narrow free line's position, quantified below.

For a target fractional error $r_*$ and chosen resolution $\eta=\alpha E$, set $q_*=[r_*/(1+r_*)]^{1/2}$. The sufficient energy window satisfies
\begin{equation}
 (1+\alpha^2)E^2+(2D-q_*\varepsilon_g\alpha)E+D^2\le0.
 \label{eq:calibrationwindow}
\end{equation}
Its endpoints are the two roots; a nonempty interval exists when $D\le q_*\varepsilon_g\alpha/[2(1+\sqrt{1+\alpha^2})]$. For $D>0$, this worst-case guarantee has a positive lower energy endpoint; $D=0$ recovers a window extending to zero. Independent fractional side-bond errors $e_1,e_2<1$ and bridge error $e_v$ are included by replacing the nominal $\varepsilon_g$ with its guaranteed lower value $\varepsilon_g(1-e_1)(1-e_2)/(1+e_v)$.

An illustrative calibration point uses $\omega=0.1J$, $\eta=0.02J$, nominal $s_1=s_2=J$, and $v=0.2J$, with a fixed target interaction $U=3J$. Equation~\eqref{eq:calibrationenvelope} gives the following guaranteed upper bounds at these parameters; the same bounds apply at every $U$:
\begin{center}
\begin{tabular}{lll}
\toprule
Side-energy budget & Bond budget & Guaranteed fractional error\\
\midrule
$D=0$ & Nominal bonds & $1.1\%$\\
$D=0.02J$ & Nominal bonds & $2.3\%$\\
$D=0.02J$ & $|v|\le0.22J$, $|s_1|,|s_2|\ge0.95J$ & $3.4\%$\\
\bottomrule
\end{tabular}
\end{center}

\subsection{Calibrating the actual free reference}

The source and the free contact structure define the spectral reference. Equations~\eqref{eq:mainbound} and \eqref{eq:positivebound} compare any two bath Hamiltonians at this fixed external structure. Target onsite shifts caused by tuning an interaction are thus already included in the bound: they are terms in $H_{B,\alpha}$. Reproducing every bath one-particle parameter is unnecessary for this comparison. Both ordinary and centered target-density interactions are allowed; their difference is an onsite term within the bath. The external contact parameters fix the reference shared by all these environments.

The common reference can also be the free source plus its first contact stage: Eq.~\eqref{eq:AC} gives $(1-\delta)\A_C\preceq\A_\alpha\preceq(1+\delta)\A_C$ for every bath. This reference is obtained by disconnecting the bridge while retaining the source and first-side parameters. It is fixed by those parameters before the bath is characterized.

Drift in the external free structure is different: it changes the reference itself. A useful conservative estimate follows from comparing two complete quadratic matrices. Let $h_{\rm act}=h_{\rm nom}+\Delta h$ be the complete quadratic matrices, with $\|\Delta h\|\le d$. The resolvent identity gives, for an embedded source vector $u$,
\begin{equation}
 \|(z-h_{\rm act})^{-1}u\|
 \le(1+d/\eta)\|(z-h_{\rm nom})^{-1}u\|.
\end{equation}
Interchanging the two matrices and using $\A_0=(\eta/\pi)P_a(z-h)^{-\dagger}(z-h)^{-1}P_a^\dagger$ yields
\begin{equation}
 (1+d/\eta)^{-2}\A_{0,\rm nom}\preceq\A_{0,\rm act}
 \preceq(1+d/\eta)^2\A_{0,\rm nom}.
 \label{eq:nominalcalibration}
\end{equation}
Equation~\eqref{eq:nominalcalibration} quantifies the sensitivity of a narrow line to uncertainty in its one-particle parameters: at $d=\eta$, its worst-case multiplicative range is $[1/4,4]$. For a uniform $1\%$ reference-intensity allowance it is sufficient to know the quadratic matrix within $d/\eta\le0.0049<\sqrt{1.01}-1$. This requirement concerns uncertainty in the calibrated matrix; the physical side offsets themselves enter Eq.~\eqref{eq:calibrationenvelope}.

These tolerances control uncertainty in the specified free reference, including source-energy drift. Changes confined to the bath are covered by the spectral comparison. A particular local probe can be less sensitive to calibration errors; its resolvent gives that sensitivity directly.

A reference measured in interleaved acquisitions uses the same contact and probe parameters as the interacting signal. The residual uncertainty between those acquisitions enters Eq.~\eqref{eq:nominalcalibration}. A fixed-interaction implementation based on specified Hubbard preparations is described in the companion.

\subsection{Fixing the common resonance by particle--hole symmetry}
\label{sec:symmetry}

The theorem allows independently chosen contact hoppings. A symmetry can additionally fix the two free side-orbital energies at a common value. Consider a real bipartite hopping network with zero onsite energies and target-only density interactions $\sum_{ij}U_{ij}(n_{r_i}-1/2)(n_{r_j}-1/2)$. Give each orbital $\alpha$ its sublattice sign $\sigma_\alpha=\pm1$. The canonical transformation $\mathcal C f_\alpha\mathcal C^{-1}=\sigma_\alpha f_\alpha^\dagger$ preserves every real hopping between opposite sublattices and reverses $n_\alpha-1/2$. The centered density interactions are therefore invariant, whereas an onsite term $\Delta_\alpha(n_\alpha-1/2)$ is odd. Exact particle--hole symmetry fixes both free side energies at zero while allowing independently chosen hopping amplitudes. Independent real bond disorder preserves this statement in each sample; uniform bounds require the side couplings to stay bounded away from zero.

This implementation uses the prescribed graph, noninteracting side orbitals, and centered target interactions. Centering is the condition for this particular symmetry implementation: an interaction written as $U n_i n_j$ requires onsite compensation to preserve particle--hole symmetry. The spectral theorem itself allows either convention without compensation. Equation~\eqref{eq:mainbound} holds for all stationary states of this Hamiltonian, including states that break particle--hole symmetry. Symmetry-controlled interference zeros have established precedents~\cite{valli2019}. The bypass in Sec.~\ref{sec:collective-bypass} illustrates the independent role of connectivity: restoring a target component on the Fermi contour also restores metallic feedback. The companion gives the corresponding spectral example and a systematic comparison of contact defects.

\subsection{A finite spectral window with weak interface interactions}
\label{sec:weak-interface-round9}

Weak source or interface interactions add a residual energy scale to the device. A particle probe can still compare with the same free reference when its linewidth exceeds that scale and remains smaller than the energy being resolved. We derive this finite window in the actual stationary state of the perturbed device, keeping the environment's size and interaction strength unrestricted.

The new interactions make the side-orbital equations nonlinear in the fermion operators. We can still use a resolvent argument by treating operators themselves as vectors: their exact time evolution is linear under the commutator with $H$. In the actual stationary state $\rho$ of the perturbed device, define
\begin{equation}
 (A,B)_\rho=\Tr\rho\{A^\dagger,B\},\qquad
 \mathcal L A=[H,A],\qquad \mathcal R_z=(z-\mathcal L)^{-1}.
 \label{eq:weak-interface-space}
\end{equation}
Operators with zero norm are identified with zero, giving a Hilbert space. In a simultaneous eigenbasis of $H$ and $\rho$, the operator $|m\rangle\langle n|$ has weight $p_m+p_n$ and real Liouvillian eigenvalue $E_m-E_n$. Thus $\mathcal L$ is self-adjoint and $\|\mathcal R_z\|\le1/\eta$. Define linear embeddings by $S_Xu=\sum_{j\in X}u_j f_j^\dagger$. Canonical anticommutation gives $S_X^\dagger S_X=I$ in every state, and $S_a^\dagger\mathcal R_zS_a=G_{aa}^\rho(z)$. Using creation operators here fixes the matrix orientation for complex Hermitian source hoppings as well as real ones.

For real uniform contact couplings, write the exact operator equations as
\begin{align}
 \mathcal L S_C&=S_Ch_C+vS_rP_b+D_C,\nonumber\\
 \mathcal L S_d&=s_2S_r+D_d,
 \qquad \|D_C\|\le k_C,\quad \|D_d\|\le k_d.
 \label{eq:weak-interface-residual}
\end{align}
Here $P_b$ selects the connector coordinates from $C=(a,b,c)$; $D_C,D_d$ contain the additional interaction terms in the equations of motion. Their norms are maps from the indicated orbital spaces into the stationary operator Hilbert space. All calibrated quadratic terms remain in the specified reference.

For a source vector $u$, set $f=g_C(z)P_a^\dagger u$. The free side equation gives $P_bf=(z/s_1)P_cf$. Two resolvent identities then give
\begin{align}
 \mathcal R_zS_au&=S_Cf+v\mathcal R_zS_rP_bf
                         +\mathcal R_zD_Cf,\nonumber\\
 s_2\mathcal R_zS_r&=z\mathcal R_zS_d-S_d-\mathcal R_zD_d.
 \label{eq:weak-interface-identities}
\end{align}
Insert the target identity into the source identity. The three resulting errors come from the first-stage residual $D_C$, propagation through the target, and the side residual $D_d$. Bounding each resolvent by $1/\eta$ gives
\begin{align}
 \|\mathcal R_zS_au-S_Cf\|&\le e(z)\|f\|,\nonumber\\
 e(z)&=\frac{k_C}{\eta}
 +\frac{|vz|}{|s_1s_2|}
 \left(\frac{|z|+k_d}{\eta}+1\right).
 \label{eq:weak-interface-error}
\end{align}
The spectral identity for a self-adjoint resolvent implies
\begin{equation}
 u^\dagger\A^\rho u=\frac{\eta}{\pi}\|\mathcal R_zS_au\|^2,
 \qquad
 u^\dagger\A_Cu=\frac{\eta}{\pi}\|f\|^2.
\end{equation}
The triangle inequality therefore proves
$(1-e)^2\A_C\preceq\A^\rho\preceq(1+e)^2\A_C$ when $e<1$.
For the complete connected quadratic reference $\A_0$, Eq.~\eqref{eq:AC} supplies
$(1-\delta)\A_C\preceq\A_0\preceq(1+\delta)\A_C$.
Combining these estimates yields the finite-interaction extension
\begin{equation}
 \boxed{
 \frac{(1-e)^2}{1+\delta}\A_0
 \preceq\A^\rho\preceq
 \frac{(1+e)^2}{1-\delta}\A_0,\qquad e<1,\quad\delta<1.}
 \label{eq:weak-interface-bound}
\end{equation}

The residual norms can be bounded locally. If every column operator $D_j$ has ordinary operator norm at most $u_*$ and overlaps the support of at most $\nu$ columns, including itself, odd parity gives $\{D_i^\dagger,D_j\}=0$ on disjoint supports. The Gram matrix then has row sum at most $2\nu u_*^2$, so
\begin{equation}
 \|D\|\le\sqrt{2\nu}\,u_*.
 \label{eq:weak-interface-local}
\end{equation}
This estimate is independent of volume and state. Finite-range Hubbard and density interactions satisfy the support condition.

At fixed contact energies and $|z|\sim E$, a sufficient small-error window is
\begin{equation}
 \max\!\left\{\frac{E^2}{\varepsilon_g},\ k_C,
                    \frac{Ek_d}{\varepsilon_g}\right\}
 \ll\eta\ll E,
 \qquad \varepsilon_g=|s_1s_2/v|.
 \label{eq:weak-interface-window}
\end{equation}
Weak source interactions thus allow a finite window above their residual energy scale. For fixed nonzero $k_C$, the guarantee applies above a finite residual scale: an allowed term $U_an_an_f$ with an occupied conserved spectator shifts a source level by $U_a$. Equation~\eqref{eq:weak-interface-window} quantifies this physical cutoff. The actual perturbed stationary state supplies the operator inner product, and the reference enters through its fixed quadratic propagator.


\subsection{Collective readout and feedback from a weak source interaction}
\label{sec:feedback-window-round10}

A residual interaction on the source has a second effect: the collective spin arriving there can act back on the low-energy metal. This feedback links the desired signal to a practical operating scale. We calculate its coefficient from the same orbital polarization that determines the signal, then find the energies above which the feedback remains weak.

Add $H_U=U_a\sum_i n_{ai\uparrow}n_{ai\downarrow}$ while keeping the predetermined quadratic reference fixed. Its spin-independent Hartree shift belongs to the residual spectral-error budget. To find the additional dynamical coupling, expand about $U_a=0$. Let $Z$ be the source weight of the low band at the Fermi contour. The source-spin matching is $M_a=Z m_\psi+b_a\phi+\cdots$, where $m_\psi$ is the low-band spin density and $b_a=2\lambda\kappa_a+O(\lambda^3)$. For small $|U_a|\nu_F$ and fixed weak $\lambda$, the induced low-band interaction is
\begin{equation}
 H_{\mathrm{fb}}=g_U\phi m_\psi,\qquad
 g_U=-U_a Z\lambda\kappa_a=-\frac{U_a Z}{2}b_a
       +O(U_a^2\lambda,U_a\lambda^3).
 \label{eq:source-feedback-round10}
\end{equation}
Here and below the expansion coefficients are evaluated at fixed, nonzero high-band gaps.

To obtain the dynamical vertex, write the free source response to a target onsite field as $\kappa_{ar}(\bm q,z)=\kappa_{ar}^{\mathrm{inter}}+\kappa_{ar}^{\mathrm{intra}}$. The interband term is analytic below the high-band gap $\Delta_h$ and approaches $\kappa_a$ in both static and dynamic limits. In the inversion-symmetric contact, its leading corrections are quadratic in frequency and momentum. The intraband term has the target vertex $e^2(e+\omega)^2$ and vanishes as the fourth power of the low-energy scale at zero temperature. The interband constant therefore survives in the low-energy response. At $q=0$ and finite frequency the free intraband term vanishes exactly, since band occupations are conserved.

Down spins see the target field $-\lambda\phi$, which induces $\delta n_{a\downarrow}=-\lambda\kappa_{ar}\phi$. The first-order Hubbard self-energy of up spins is $U_a\delta n_{a\downarrow}$; the down-spin term has the opposite sign. Projecting the external source density onto the low band supplies $Z$ and gives Eq.~\eqref{eq:source-feedback-round10}. An opposite-spin onsite interaction has no Fock contraction at this order. Equivalently, the cross term in $(U_a/4)(n_a^2-M_a^2)$ gives the same coefficient. Cancellation of the summed orbital polarization, $\sum_\alpha b_\alpha=0$, expresses total-spin conservation. For a Hubbard interaction acting on $a$ alone, the feedback is set by that orbital's polarization.

Changes of the spin-symmetric background affect this leading vertex at order $U_a^2\lambda$. The induced vertex restores ordinary metallic damping at sufficiently low energy. For a circular Fermi surface, with density of states per spin $\nu_F$,
\begin{equation}
 \Pi_U''(q,\omega)=2\nu_F g_U^2
 \frac{\omega}{\sqrt{(v_Fq)^2-\omega^2}},
 \qquad 0<\omega<v_Fq.
 \label{eq:feedback-damping-round10}
\end{equation}
Its coefficient is tied to the measured collective overlap:
$2\nu_Fg_U^2=(\nu_FU_a^2Z^2/2)b_a^2$ to leading order. The damping concerns small nonzero momentum; conservation still removes finite-frequency intraband absorption at exactly $q=0$.

\paragraph{Compare feedback with the measured collective response.}
The damping coefficient alone does not tell us when feedback becomes important: it must be compared with the original collective response. Let $S=b_a\phi$ be the collective part of the source spin and $X_a=\chi_{aa,0}^{\rm coll}=b_a^2\chi_{\phi\phi}$ its response at zero residual source interaction. Then $H_{\rm fb}=-(U_a Z/2)S m_\psi$. After the static feedback is included in the tuned critical mass, integrating out the low band adds a dynamical kernel proportional to $\Pi_{\rm dyn}=\chi_\psi(q,0)-\chi_\psi(q,\ii\Omega)$. The dimensionless comparison is
\begin{equation}
 \mathfrak f(q,\ii\Omega)=\frac{U_a^2Z^2}{4}
 \Pi_{\rm dyn}(q,\ii\Omega)X_a(q,\ii\Omega),\qquad
 \Pi_{\rm dyn}=\frac{2\nu_F|\Omega|}{\sqrt{\Omega^2+(v_Fq)^2}}.
 \label{eq:observable-feedback-round11}
\end{equation}
The second expression uses a circular Fermi surface. The product compares the added kernel with the inverse collective response: $\mathfrak f\ll1$ selects weak feedback, while $\mathfrak f=1$ estimates the departure scale. Because $X_a$ is the physical source response, this comparison is independent of how the Ising field is normalized.

To find that scale, insert the ideal critical response
$\chi_{\phi\phi}(q,\ii\Omega)=\mathcal C_\phi(\Omega^2+c_{\rm I}^2q^2)^{-1+\eta_{\rm I}/2}$.
Follow a fixed ratio $|\Omega|/q=u$ and write $E=(\Omega^2+c_{\rm I}^2q^2)^{1/2}$. Solving $\mathfrak f=1$ gives
\begin{equation}
 E_U^{2-\eta_{\rm I}}
 =\frac{\nu_FU_a^2Z^2}{2}
   [b_a^2\mathcal C_\phi]\,
   \frac{u}{\sqrt{v_F^2+u^2}}.
 \label{eq:feedback-scale-round10}
\end{equation}
The bracket is the singular source-response amplitude, so measuring it fixes the contact factor in the scale estimate. At fixed $\lambda$, the departure scales as $E_U\propto|U_a|^{2/(2-\eta_{\rm I})}$. This is the leading scale comparison above any additional residual-interaction instability. Connected Ising correlations, starting with the four-point function in the perturbative response, determine the full crossover function and its numerical coefficients. The full source response also contains the regular metallic term in $M_a=Zm_\psi+S+\cdots$.

A generated coupling to the Ising energy density starts at $U_a\lambda^2$ and produces a departure with power $2/(3-2\Delta_\epsilon)$ in $|U_a|$. Its scale is asymptotically lower when the order-parameter vertex in Eq.~\eqref{eq:source-feedback-round10} is nonzero. The order-parameter feedback therefore sets the first collective restriction used below.

\subsection{Temperature and the common measurement window}
\label{sec:thermal-window-round10}

At nonzero temperature, electrons occupy a range of energies around the target node. Their small but finite target weights restore a correspondingly small damping channel. We evaluate this thermal correction and combine it with the residual-interaction and spectral-resolution scales. For a linearized Fermi surface with constant density of states $\nu_F$ and target weight $|u_r(e)|^2=C_r e^4$, the absorption phase space is
\begin{align}
 I_4(\omega,T)&=\int_{-\infty}^{\infty}\dd e\,
 [f_T(e)-f_T(e+\omega)]e^4(e+\omega)^4\nonumber\\
 &=\frac{\omega^9}{630}
 +\frac{124\pi^6}{63}\omega^3T^6
 +\frac{127\pi^8}{15}\omega T^8,
 \label{eq:thermal-polynomial-round10}
\end{align}
where $f_T(e)=(e^{e/T}+1)^{-1}$ and $\omega>0$. To derive it, use $f_T(e)-f_T(e+\omega)=\int_0^\omega[-f_T'(e+s)]\dd s$ and set $x=e+s$. The even moments of $-f_T'(x)$ are
\begin{equation}
 \mu_0=1,\quad\mu_2=\frac{\pi^2T^2}{3},\quad
 \mu_4=\frac{7\pi^4T^4}{15},\quad
 \mu_6=\frac{31\pi^6T^6}{21},\quad
 \mu_8=\frac{127\pi^8T^8}{15}.
\end{equation}
The shift integrals multiplying $\mu_2$ and $\mu_4$ vanish; the remaining terms are $\omega^9/630+(4\omega^3/3)\mu_6+\omega\mu_8$. The missing $T^2$ and $T^4$ corrections reflect the fourth-order zeros at both ends of the zero-temperature absorption interval.

The leading damping is
\begin{equation}
 \Pi_{\mathrm{node}}''(q,\omega,T)
 =\frac{2\lambda^2\nu_FC_r^2 I_4(\omega,T)}
 {\sqrt{(v_Fq)^2-\omega^2}}.
 \label{eq:thermal-damping-round10}
\end{equation}
For $T\ll\omega$ it approaches the zero-temperature result; for $\omega\ll T$ it has ordinary $\omega/q$ kinematics with a coefficient proportional to $T^8$. Within this leading approximation the relative thermal correction is
\begin{equation}
 \frac{I_4(\omega,T)}{I_4(\omega,0)}-1
 =1240\pi^6(T/\omega)^6+5334\pi^8(T/\omega)^8.
\end{equation}
It is below $10\%$ for $T/\omega\le0.064$. These expressions apply when $T$, $\omega$, and $v_Fq$ are below the band-curvature and interband scales. At energies below $T$, the spin sector also crosses into its thermal regime.

A common finite window remains with weak source interactions. Let $E_L$ denote the relevant finite-size energy, $E_r$ the departure scale from critical-mass detuning, and $E_{\mathrm{geom}}$ the feedback scale from imperfect contacts. Set $\Lambda=\min(\Delta_h,E_{\mathrm{lin}})$, where $E_{\mathrm{lin}}$ is the low-band Fermi-surface linearization scale; this can be much smaller than the interband gap. Sufficient conditions for comparing the two responses are
\begin{align}
 \max\{T,E_L,E_r,E_U,E_{\mathrm{geom}}\}&\ll E\ll\Lambda,
 \label{eq:joint-collective-window-round10}\\
 \max\{E^2/\varepsilon_g,k_C,Ek_d/\varepsilon_g\}
 &\ll\eta_{\mathrm{sp}}\ll E.
 \label{eq:joint-spectrum-window-round10}
\end{align}
The residual norms $k_C,k_d$ are defined by the interface equations of motion; the spectral and collective widths can be chosen independently. For an uncentered onsite source Hubbard term alone, $k_C\le|U_a|$. Indeed, the local residual creation map is
$D(v)=U_a(v_\uparrow a_\uparrow^\dagger n_\downarrow+v_\downarrow a_\downarrow^\dagger n_\uparrow)$.
It maps the one-particle sector into the doubly occupied state and obeys $\|\{D(v)^\dagger,D(v)\}\|\le U_a^2\|v\|^2$. Residuals on distinct sites have vanishing odd anticommutators, giving the same bound independently of volume and state.

For fixed weak $\lambda$ and finite matching constants, $E_U/|U_a|\propto|U_a|^{\eta_{\rm I}/(2-\eta_{\rm I})}\to0$. Hence the conservative spectral requirement $|U_a|\ll\eta_{\mathrm{sp}}\ll E$ eventually exceeds the feedback cutoff and also covers the Hartree drift relative to the fixed reference. For example, with $u=|U_a|/\Lambda\to0$, the choice $E=\Lambda u^{1/2}$ and $\eta_{\mathrm{sp}}=\Lambda u^{3/4}$ makes both the ideal contact error and the residual spectral error vanish, while $E_U/E\to0$. Temperature, finite size, and critical detuning must separately satisfy Eq.~\eqref{eq:joint-collective-window-round10}. Tuning the Ising mass to its interacting critical value is a phase-control step; the spectral reference remains unchanged. The same virtual polarization therefore fixes both the collective signal and its residual feedback within a controlled joint window.



\bibliographystyle{unsrtnat}
\bibliography{references}